\documentclass[10pt]{book}
\usepackage{array}
\usepackage{latexsym, amsbsy, amssymb, amsmath, bm, bbm}
\usepackage{graphicx}
\RequirePackage[OT1]{fontenc} \RequirePackage{amsthm,amsmath}
\RequirePackage[numbers]{natbib}
\usepackage{color}
\usepackage{multirow}

\def\bc{\begin{center}}
\def\br{\begin{array}}
\def\er{\end{array}}
\def\ec{\end{center}}
\def\bmp{\begin{minipage}}
\def\emp{\end{minipage}}
\def\beq{\begin{equation}}              
\def\eeq{\end{equation}}

\def\bfbeta{\mbox{\boldmath $\beta$}}

\def\bc{\begin{center}}                 
\def\ec{\end{center}}                   
\def\beq{\begin{equation}}              
\def\eeq{\end{equation}}                
\def\br{\begin{array}}                  
\def\er{\end{array}}                    
\def\beqr{\begin{eqnarray}}             
\def\eeqr{\end{eqnarray}}               
\def\beqrs{\begin{eqnarray*}}           
\def\eeqrs{\end{eqnarray*}}     

\newtheorem{theorem}{\bf Theorem}
\newtheorem{assumption}{\bf Assumption}
\newtheorem{definition}{\bf Definition}

\newtheorem{lemma}{\bf Lemma}

\newtheorem{remark}{\bf Remark}
\renewcommand{\thefootnote}{\arabic{footnote}}

\DeclareMathOperator*{\argmin}{arg\,min}
\DeclareMathOperator{\sign}{sign}
\def\bmx{\bm x}
\def\bmX{\bm X}

\def\bmv{\bm v}
\def\bmu{\bm u}
\def\bmbeta{\bm \beta}
\def\bmtheta{\bm \theta}
\def\bmgamma{\bm \gamma}

\def\bmomega{\bm \omega}
\def\bmxi{\bm \xi}
\def\bmdelta{\bm \delta}
\def\bmDelta{\bm \Delta}
\def\bmSigma{\bm \Sigma}
\def\bmzero{\bm 0}
\def\what{\widehat}
\def\wtilde{\widetilde}

\def\yij{y_{i\backslash j}}
\def\bmxij{\bmx_{i\backslash j}}
\newcommand{\matr}[1]{\mathbf{#1}} 
\def\matrM{\matr{M}}
\def\matrW{\matr{W}}

\newcommand{\E}{\mathrm{E}}

\newcommand{\Cov}{\mathrm{Cov}}

\begin{document}


\renewcommand{\baselinestretch}{1.2}

\markright{
\hbox{\footnotesize\rm Statistica Sinica (2017): Preprint}\hfill
}

\markboth{\hfill{\footnotesize\rm JIWEI ZHAO, YANG YANG AND YANG NING} \hfill}
{\hfill {\footnotesize\rm VARIABLE SELECTION WITH NONIGNORABLE MISSING DATA} \hfill}

\renewcommand{\thefootnote}{}
$\ $\par


\fontsize{10.95}{14pt plus.8pt minus .6pt}\selectfont
\vspace{0.8pc}
\centerline{\large\bf Penalized pairwise pseudo likelihood for variable selection}
\vspace{2pt}
\centerline{\large\bf with nonignorable missing data}
\vspace{.4cm}
\centerline{Jiwei Zhao$^{1}$, Yang Yang$^{1}$ and Yang Ning$^{2}$}
\vspace{.4cm}
\centerline{\it $^{1}$State University of New York at Buffalo, $^{2}$Cornell University}
\vspace{.55cm}
\fontsize{9}{11.5pt plus.8pt minus .6pt}\selectfont

\vspace{-3mm}
\begin{quotation}
\noindent {\it Abstract:}
The regularization approach for variable selection was well developed for a completely observed data set in the past two decades. In the presence of missing values, this approach needs to be tailored to different missing data mechanisms. In this paper, we focus on a flexible and generally applicable missing data mechanism, which contains both ignorable and nonignorable missing data mechanism assumptions. We show how the regularization approach for variable selection can be adapted to the situation under this missing data mechanism. The computational and theoretical properties for variable selection consistency are established. The proposed method is further illustrated by comprehensive simulation studies and real data analyses.\par

\vspace{9pt}
\par
\end{quotation}\par


\fontsize{10.95}{14pt plus.8pt minus .6pt}\selectfont
\vspace{-2mm}

\section{Introduction}

Variable selection is an important topic in regression analysis. In the past two decades, researchers investigated a series of regularization approaches for variable selection and developed both theoretical and computational properties.
The two mainstream techniques are the LASSO (Least Absolute Shrinkage and Selection Operator; \cite{tibshirani1996regression}), or the $L_1$-penalization, and the nonconvex penalizations such as the SCAD (Smoothly Clipped Absolute Deviation; \cite{fan2001variable}) and the MCP (Minimax Concave Penalty; \cite{zhang2010nearly}).
The LASSO owns its popularity largely due to its computational convenience, but it induces estimation bias for parameters with large absolute values.
Using a nonconvex penalty, one has to minimize a nonconvex function, which raises extra computational challenges, but the intrinsic estimation bias of the LASSO can be eliminated and corrected.
With completely observed data, it's shown that, under regularity conditions, both of the two mainstream techniques achieve variable selection consistency properties; however, when missing values are present the appropriate regularization approaches are relatively limited in the literature. In this paper, we explore how to use both techniques for variable selection in the presence of missing data.

In the missing data literature, one often defines an indicator $R$ to illustrate whether the data from each subject are completely available or not, i.e., $R=1$ represents the completely observed subject, and $R=0$ otherwise.
The probability distribution function of $R$ conditional on all the data, termed as the missing data mechanism \cite{little2002statistical}, should be incorporated in the analysis compensating for the effect of missing data.
There are various missing data mechanism assumptions.
Briefly, if it only depends on the completely observed data, the mechanism is called missing at random (MAR); otherwise, it is called missing not at random, or nonignorable.
The likelihood-based methods, usually under the MAR assumption, can be derived for variable selection. Most of the currently existing literature falls in this category. For example, \cite{ibrahim2008model} developed likelihood methods for the computation of model selection criteria based on the output of the EM algorithm. They derived a class of information criteria for missing data problems. \cite{garcia2010variable} considered the regularization approach using SCAD or adaptive LASSO and adopted the EM technique to formulate the observed likelihood for variable selection in a low-dimensional setting.

In general, likelihood-based methods need to specify a parametric distribution of the missing data mechanism. One has to be cautious about this type of assumptions.
First, it is well known that a parametric assumption is very sensitive and may easily induce a misspecified model. If this happens, it prompts biased estimation and inaccurate selection results.
Second, although MAR occurs in some applications, in many situations there is a suspicion that the missing data mechanism is nonignorable \cite{ibrahim1999missing}. For nonignorable missing data, applying methods derived under the MAR assumption may result in serious estimation bias and incorrect conclusion.
Third, the situation with nonignorable missing data is generally more challenging to deal with. One notorious feature of nonignorable missingness is the identifiability issue \cite{robins1997toward}. In theory, one has to first carefully study the model identification conditions before doing any statistical analyses. The readers may refer \cite{kim2013statistical} for the most recent development of nonignorable missing data.

Due to the complexity of the missing data mechanism assumptions, in reality one may carry out the sensitivity analysis to validate the analysis results. The other preferred and ideal remedy is to impose an assumption as flexible and generally applicable as possible. This type of assumption usually does not specify a parametric model, and often, it is called unspecified missing data mechanism. The work of \cite{liang2000regression, tang2003analysis, shao2013estimation, zhao2015semiparametric, fang2017likelihood, zhao2017reducing} follows this direction. In their work, usually a nonregular likelihood can be derived and will be the base for the following inference. For instance, based on the idea of conditional likelihood \cite{kalbfleisch1978likelihood}, \cite{liang2000regression}
introduced a separable/decomposable missing data mechanism under a generalized linear model framework and developed a pairwise pseudo likelihood to find an estimator of the unknown parameter. Since their missing data mechanism is unspecified and contains many nonignorable scenarios, not all of the unknown parameters are estimable due to the non-identification issue.

In this paper, our motivation is to conduct variable selection with missing data, more interestingly, with nonignorable missing data.
This paper contains three major novel contributions.
First, under the high dimensional setting, we consider the generalized linear model (GLM; \cite{mccullagh1989generalized}), which can be applied to either continuous or categorical data, and is very popular in real applications. We impose an unspecified missing data mechanism assumption, which is flexible and generally applicable and robust for the potential model misspecification. Besides MAR cases, it contains many nonignorable scenarios. Under this assumption, although not all parameters are identifiable, a pseudo likelihood function which produces an estimator of a dispersion-scaled version of the original parameter is developed. We show that, the variable selection can be carried out by penalizing the aforementioned pseudo likelihood through the estimable dispersion-scaled parameter. Here the fact is that, due to the messy missing data and the flexible mechanism assumption, we may not fully retrieve all the information contained in the original data, hence we are not able to estimate all the unknown parameters. However, the key idea is that, our regularization procedure can still be carried out for the purpose of variable selection, based on the pseudo likelihood function and the estimable dispersion-scaled parameter.

Second, numerically we propose algorithms to efficiently optimize the penalized pseudo likelihood for both the LASSO and nonconvex penalties. This is not a trivial task due to the complicated U-statistic structure in the pseudo likelihood function. For the LASSO penalty, we find that, the objective function can be transformed to the penalized likelihood function for a standard penalized logistic regression model without the intercept term after some data manipulation. More importantly, for the nonconvex penalties, we developed an iterative algorithm based on the trick we used in the LASSO and the local linear approximation (LLA; \cite{zou2008one, fan2014strong}).

Third, theoretically we show that, under the high dimensional setting, the variable selection consistency can be achieved with some mild regularity conditions. The challenges are not only from the complicated pairwise U-statistic structure in the pseudo likelihood, but also from the nonconvex penalties that we devote to. As we introduced in the first paragraph, the theoretical arguments for the nonconvex penalties are generally more sophisticated than those for the LASSO penalty. We first develop some concentration inequalities for a general U-statistic. To perform variable selection, we define the oracle estimator and provide a sharp bound to control the rate of the oracle estimator. Finally we show that the support set of our proposed estimator is the same as that of the oracle estimator, and hence the selection consistency is achieved. With respect to the high dimensional set-up, our theory allows the dimensionality to grow at most as fast as an exponential rate of the sample size. The details can be found in Section 4.

Indeed there are scarce literature on high dimensional problems with missing data.
\cite{loh2012high} considered a linear model with covariates may have missing values and studied the theoretical properties of the estimators using a regularization approach via the LASSO. Their results only apply under the most simplest missing data mechanism assumption: missing completely at random. It is well known that this assumption is too restrictive and may not be realistic.
More recently, \cite{ning2017likelihood} showed some results on parameter estimation in a similar context, however, our paper is distinctively different from them in every aspect. In terms of our first contribution, \cite{ning2017likelihood} only studied the parameter estimation problem, however, their procedure cannot estimate the dispersion parameter. In contrast, we devote ourselves to the variable selection problem and as we described above, we show that the regularization procedure can still be carried out through the dispersion-scaled version of the original parameter. In terms of our second contribution, \cite{ning2017likelihood} only studied the LASSO penalty while we also propose algorithms to efficiently optimize the penalized pseudo likelihood for the nonconvex penalties. In terms of our third contribution, \cite{ning2017likelihood} considered the estimation problem and an associated inference procedure with the LASSO penalty, which is very different from ours.
In this paper, we devote ourselves to the investigation of variable selection and many of our theoretical techniques and tools are distinct from them.
For example, to perform variable selection, we require the minimal signal strength condition which is generally not needed for parameter estimation. In addition, unlike the theoretical analysis for parameter estimation, we define the oracle estimator and provide a sharp bound to control the rate of the oracle estimator.

The remainder of the paper is organized as follows. In Section 2, we first provide a brief review of the regularization approach in the case of no missing data, and then introduce our proposed penalized pseudo likelihood. The algorithms designed for both the LASSO and nonconvex penalties are presented in Section 3. Section 4 contains the theoretical results on variable selection consistency, Section 5 includes the numerical results illustrating the finite sample performance of our proposed method and its comparison with some existing methods and Section 6 provides two real data analyses. In Section 7, we conclude our paper with a discussion. All the technical details are retained to the Appendix.

\section{Methodology}

\subsection{Brief Review in the Case of No Missing Data}

Assume that we have a collection of independent observations $\{y_i, \bmx_i\}, i=1,\ldots,N$, where $(y_i, \bmx_i)$'s are identically distributed realizations of $(Y,\bmX)$. We let $Y$ denote the scalar response variable, and $\bmX$ be a $p$-dimensional covariate variable. Assume that, with a canonical link, the conditional distribution of $Y$ given $\bmX$ belongs to a generalized linear model (GLM; \cite{mccullagh1989generalized}) with the following density:
\begin{eqnarray}
p(Y|\bmX;\bmtheta) = \exp[\phi^{-1}\{Y\eta-b(\eta)\} + c(y;\phi)],
\label{glm}
\end{eqnarray}
where $b$ and $c$ are known functions, $\eta=\alpha+\bmbeta^T \bmX$, $\bmtheta = (\alpha, \bmbeta^T, \phi)^T$, $\phi$ represents the positive dispersion parameter.

To carry out variable selection through the regularization approach, we target to obtain the minimizer of the following penalized likelihood function
\begin{eqnarray}
-\frac1N \sum_{i=1}^N \log p(y_i|\bmx_i;\bmtheta) + \sum_{j=1}^p p_\lambda(|\beta_j|),
\label{plikewom}
\end{eqnarray}
where $p_\lambda(t)$ represents a penalty function, and $\lambda\geq 0$ is the tuning parameter. Note that the penalty term is only applied to $\bmbeta$. The variable selection can be achieved without estimating the dispersion parameter $\phi$.

In the LASSO penalty, or the $L_1$-penalty, $p_\lambda (t) = \lambda |t|$, and $p'_\lambda(t) = \lambda$ for $t > 0$. Due to the convexity of the LASSO, the coordinate descent algorithm \cite{friedman2010regularization} has been shown to be very efficient to minimize (\ref{plikewom}).
Theoretically, a strong irrepresentable condition is necessary for the LASSO to be selection consistent \cite{zhao2006model}. Also, the LASSO may induce intrinsic estimation bias for parameters with large absolute values.

In \cite{fan2001variable}, the authors advocated penalty functions that provide estimators with three properties: sparsity, unbiasedness and continuity. Clearly, the $L_1$-penalty does not satisfy the unbiasedness property. In this class of nonconvex penalty functions, two frequently used representatives are the SCAD and the MCP. The SCAD is due to \cite{fan2001variable}:
\[
p'_\lambda(t) = \lambda \mathbbm{1}(t\leq \lambda) + \frac{(a\lambda-t)_{+}}{a-1} \mathbbm{1}(t> \lambda),
\]
for some $a>2$ and $t>0$, where $\mathbbm{1}(\cdot)$ is the indicator function;
the MCP is due to \cite{zhang2010nearly}:
\[
p'_\lambda(t) = \frac{(a\lambda - t)_{+}}{a},
\]
for some $a>0$ and $t>0$. Numerous papers have been devoted to  study the statistical properties of the resulting estimators, for instance, \cite{fan2010selective, fan2011nonconcave} and the references therein. However, the computation for this approach is much more involved, because the resulting optimization problem (\ref{plikewom}) is nonconvex and may have multiple local minimizers. \cite{fan2001variable} proposed the local quadratic approximation algorithm as a unified method for optimizing the nonconvex penalized likelihood; while \cite{zou2008one} worked out the local linear approximation (LLA) algorithm which turns a nonconvex penalization problem into a series of reweighed $L_1$-penalization problems. Both of them are relevant to the majorization-minimization (MM) principle \cite{hunter2004tutorial}.

\subsection{Variable Selection with Missing Data}

When variables from $(Y,\bmX)$ have missing values, recall that we have the indicator $R$ illustrating whether the data from each subject are completely observed. Without loss of generality, we assume the first $n$ subjects are fully observed with $r_i=1$, $i=1,\ldots,n$, and the remaining $N-n$ subjects may contain missing components with $r_i=0, i=n+1, \ldots, N$.

The foremost difficulty dealing with missing data is the assumption on the missing data mechanism, i.e, $\Pr(R=1|Y,\bmX)$. The most straightforward way is to impose a parametric model on it. Then the likelihood based methods can be developed for model selection, especially for the MAR case \cite{ibrahim2008model, garcia2010variable}. However, there are a few limitations to adopt this approach. First, a parametric assumption is very sensitive and may easily incur the model misspecification issue. This could happen within MAR situations, or within nonignorable situations, or between. If an incorrect parametric assumption is imposed, neither estimation nor selection results are reliable. Second, in many applications, it is plausible that the mechanism is nonignorable missingness, which is certainly a more challenging problem than the MAR case. Since the underlying truth of the missing data mechanism is unknown and its assumption is unverifiable, 
it is ideal to impose an assumption, which is more robust than a single parametric model, and is as flexible and generally applicable as possible.

Therefore, in this paper, we impose a general assumption
\begin{eqnarray}
  \Pr(R=1|Y,\bmX) = s(Y)t(\bmX),
  \label{assume}
\end{eqnarray}
where $s$ and $t$ are some functions, not necessarily to be known or specified.
As we can see, (\ref{assume}) only assumes that $\Pr(R=1|Y,\bmX)$ can be written as the multiplier of an $\bmX$-only function and a $Y$-only function. We do not impose any concrete form on $s$ or $t$, therefore, it is robust to misspecification of $s$ or $t$ function. We assume $0<\Pr(R=1)<1$ throughout.

This assumption is very flexible and it includes many specific scenarios commonly seen in the missing data literature. For example, if we consider the case of $Y$ having missing values and $\bmX$ fully observed, i.e., the missing response case, the MAR assumption studied in \cite{qin2008efficient} and among many others, belongs to a special case of (\ref{assume}) if we let $s$=constant; the nonignorable nonresponse assumption in \cite{tang2003analysis} also belongs to a special case of (\ref{assume}) if we let $t$=constant.
Besides, the situations included in (\ref{assume}) can also allow the covariate $\bmX$ to have missing values, and both response $Y$ and covariate $\bmX$ to have missing values.

Due to its flexibility, the assumption (\ref{assume}) was explored in the literature in different aspects. \cite{chan2013nuisance} considered the problem of nuisance parameter elimination in a proportional likelihood ratio model under this assumption. \cite{zhao2017approximate} studied the identifiability conditions in a GLM with non-canonical link under this assumption. Both of them only considered the classic low-dimensional statistical models. More recently, \cite{ning2017likelihood} studied the parameter estimation problem and an associated inference procedure under this assumption in a high-dimensional setting. In this paper, we focus on the variable selection problem in a high-dimensional GLM. Although under the same assumption, our work has to handle extraordinary challenges due to the high dimensionality, compared to \cite{chan2013nuisance} and \cite{zhao2017approximate}. Although it is also under the high-dimensional framework, \cite{ning2017likelihood} mainly addressed the estimation problem, which, as we described in our Introduction, needs different analytic tools than our paper.


Because of the complexity of the missing data structure and the presence of unknown functions $s$ and $t$, we propose the following pseudo likelihood function. Note that
\begin{eqnarray}
  p(Y|\bmX,R=1)=\frac{\Pr(R=1|Y,\bmX)}{w(\bmX)} p(Y|\bmX),
  \label{fact}
\end{eqnarray}
where $w(\bmX)=\int \Pr(R=1|Y,\bmX) p(Y|\bmX) dY = \Pr(R=1|\bmX)$.
Under the separable missing data mechanism assumption (\ref{assume}), $\Pr(R=1|Y,\bmX)/w(\bmX)$ in (\ref{fact}) preserves to be the multiplier of an $\bmX$-only function $s(\bmX)/w(\bmX)$ and a $Y$-only function $t(Y)$. Therefore, restricting attention to completely observed subjects with subscripts ranging from $\{1,\ldots,n\}$, decomposing $\{y_1, \ldots, y_n\}$ as rank statistics and order statistics, and conditioning on the order statistics $\{y_{(1)}, \ldots, y_{(n)}\}$, we have the following conditional likelihood for $\bmtheta$:
\begin{eqnarray*}
  p(y_1, \ldots, y_n|r_1=\ldots=r_n=1, \bmx_1, \ldots, \bmx_n, y_{(1)}, \ldots, y_{(n)}).
\end{eqnarray*}
After some derivations, it can be shown that this conditional likelihood equals
\begin{eqnarray}
  \frac{\Pi_{i=1}^n p(y_i|\bmx_i;\bmtheta)}{\sum_c \Pi_{i=1}^n p(y_{(i)}|\bmx_i;\bmtheta)},
  \label{cl}
\end{eqnarray}
where the summation in the denominator corresponds to all possible permutations of $\{1,\ldots,n\}$.

A nice feature of this method is that, it is now nuisance free: all $s$, $t$ and $w$ functions are all canceled out through conditioning.
This idea was first outlined in \cite{kalbfleisch1978likelihood}, but in practice, (\ref{cl}) encounters a tremendous computational burden with an order of $n!$ \citep{liang2000regression}. To reduce the computational burden, \citep{liang2000regression} advocated the following pairwise pseudo likelihood
\begin{eqnarray}
  \prod_{1\leq i<j\leq n} \frac{p(y_i|\bmx_i;\bmtheta)p(y_j|\bmx_j;\bmtheta)}{p(y_i|\bmx_i;\bmtheta)p(y_j|\bmx_j;\bmtheta)+p(y_i|\bmx_j;\bmtheta)p(y_j|\bmx_i;\bmtheta)}.
  \label{pacl}
\end{eqnarray}

Under the GLM assumption, the negative part of the log-version of (\ref{pacl}), after adding a normalizing constant, can be written as
\begin{eqnarray}
  \mathcal{L}(\bmgamma) = \frac{2}{n(n-1)}\sum_{1\leq i<j\leq n}\log\{1+\exp(-\yij\bmxij^T\bmgamma)\},
  \label{acloglike}
\end{eqnarray}
where $\yij = y_i - y_j$, $\bmxij = \bmx_i - \bmx_j$ and $\bmgamma=\bmbeta/\phi$.
To perform variable selection, we propose to minimize the penalized pairwise pseudo likelihood
\begin{eqnarray}
  \mathcal{L}(\bmgamma) + \sum_{j=1}^p p_\lambda(|\gamma_j|) = \frac{2}{n(n-1)}\sum_{1\leq i<j\leq n}\log\{1+\exp(-\yij\bmxij^T\bmgamma)\} + \sum_{j=1}^p p_\lambda(|\gamma_j|),
  \label{plike}
\end{eqnarray}
and we denote the minimizer as $\what\bmgamma$.
It can be seen that, the unpenalized component $\mathcal{L}(\bmgamma)$ is a U-statistic, where even the original function $b$ in the definition of GLM disappears. Since our method is under a very flexible and generally applicable assumption (\ref{assume}), to compensate for missing data, not surprisingly, we may not estimate the whole unknown parameter $\bmtheta$ itself. Instead, we can only estimate a dispersion-scaled parameter $\bmgamma=\bmbeta/\phi$ and we carry out variable selection through this dispersion-scaled parameter.

In the following, we will provide both computational and theoretical properties for variable selection through the regularization approach (\ref{plike}).

\section{Computational Algorithms}

Note that the unpenalized component $\mathcal{L}(\bmgamma)$ in (\ref{plike}) is a U-statistic and it is not trivial to be optimized. In this Section, we propose tractable and efficient algorithms to minimize (\ref{plike}) for the LASSO and nonconvex penalties respectively. We also discuss how to choose the regularization tuning parameter $\lambda$ in this Section.

\subsection{Algorithm for the LASSO}

The unpenalized component $\mathcal{L}(\bmgamma)$ in (\ref{plike}) can be written as
\begin{eqnarray*}
\mathcal{L}(\bmgamma) &=& \frac{2}{n(n-1)}\sum_{1\leq i<j\leq n}\log\{1+\exp(-\yij\bmxij^T\bmgamma)\}\\
&=& \frac{2}{n(n-1)}\sum_{1\leq i<j\leq n}\log\{1+\exp(-\sign(\yij)|\yij|\bmxij^T\bmgamma)\}\\
&=& \frac{2m}{n(n-1)} \cdot \frac{1}{m}\sum_{k=1}^m \log\{1+\exp(w_k \bmv_k^T\bmgamma)\} + \left\{1-\frac{2m}{n(n-1)}\right\}\log(2),
\end{eqnarray*}
where we let $m$ denote the number of terms in the summation across $1\leq i<j\leq n$ such that $\yij\neq 0$. For example, when
$Y$ is continuous, $m=n(n-1)/2$; when $Y$ is binary, $m=n_0 n_1$, where $n_0$ is the total number of 0's and $n_1$ is the total number of 1's, and $n_0+n_1=n$. Also, we let $\sign(\cdot)$ denote the sign function,
and we define $w_k = -\sign(\yij)$ and $\bmv_k = \bmxij|\yij|$ for $k=1, \ldots, m$.

It can be seen that, the essential component $\frac{1}{m}\sum_{k=1}^m \log\{1+\exp(w_k \bmv_k^T\bmgamma)\}$ in $\mathcal{L}(\bmgamma)$ can be treated as the negative log-likelihood function of a regular logistic regression with response $u_k$, covariate $\bmv_k$, without the intercept term, where
\[
u_k = \begin{cases}
  1 & \mbox{if } \yij>0\\
  0 & \mbox{if } \yij<0.
\end{cases}
\]

Therefore, to minimize (\ref{plike}) with the LASSO penalty, after the aforementioned data manipulation, it can be carried out directly as a regular penalized logistic regression forcing the intercept to zero, with in total $m$ subjects where the $k$-th subject has response $u_k$ and covariate $\bmv_k$.
In R, this procedure can be implemented using the package \verb"glmnet" \cite{friedman2010regularization}.


\subsection{Algorithm for Nonconvex Penalties}

With nonconvex penalties such as the SCAD and the MCP, we adopt the similar data manipulation technique as for the LASSO, and the LLA algorithm \cite{zou2008one, fan2014strong}. The LLA algorithm transforms a concave regularization problem into a series of weighted $L_1$-penalization problems by taking advantage of the nonconvex structure of the penalty functions and the MM principle. Moreover, the MM principle has provided theoretical guarantee on the convergence of the LLA algorithm to a stationary point of the nonconvex penalization problem. In \cite{fan2014strong}, the authors showed that, as long as the problem is localizable and the oracle estimator is well behaved, one can obtain the oracle estimator by using the one-step LLA. In addition, once the oracle estimator is obtained, the LLA algorithm converges, i.e., it produces the same estimator in its following iterations. Here, we summarize the details of the LLA algorithm as follows:
\begin{itemize}
  \item[1.] Initialize $\what\bmgamma^{(0)}=(\hat\gamma_1^{(0)}, \ldots, \hat\gamma_p^{(0)})^T$ and compute the adaptive weight
  \[
  \what\bmomega^{(0)} = (\hat\omega_1^{(0)}, \ldots, \hat\omega_p^{(0)})^T = (p'_\lambda(|\hat\gamma_1^{(0)}|), \ldots, p'_\lambda(|\hat\gamma_p^{(0)}|))^T.
  \]
  \item[2.] For $m=1,2,\ldots$, repeat the LLA iteration till convergence
  \begin{itemize}
    \item[2.a] Obtain $\what\bmgamma^{(m)}$ by solving the following optimization problem
    \begin{eqnarray}
    \what\bmgamma^{(m)} = \argmin_{\bmgamma} \left\{\mathcal{L}(\bmgamma) + \sum_{j=1}^p \hat\omega_j^{(m-1)} |\gamma_j|\right\},
    \label{lla}
    \end{eqnarray}
    \item[2.b] Update the adaptive weight vector $\what\bmomega^{(m)}$ with $\hat\omega_j^{(m)}=p'_\lambda(|\hat\gamma_j^{(m)}|)$.
  \end{itemize}
\end{itemize}

In our numerical studies, the initial $\what\bmgamma^{(0)}$ is chosen as the LASSO solution. In R, the major step (\ref{lla}) is implemented using the package \verb"glmnet".

\subsection{Tuning Parameter Selection}

How to select the regularization parameter $\lambda$ is of paramount importance in penalized likelihood estimation since $\lambda$ governs the complexity of the selected model. A large value of $\lambda$ tends to choose a simple model, whereas a small value of $\lambda$ inclines to a complex model. The trade-off between the model complexity and the prediction accuracy yields an optimal choice of $\lambda$. This is frequently done by using a $K$-fold cross-validation. Specifically, we denote the data set indexed by $\{1,\ldots,n\}$ as $T$, and cross validation training and test sets by $T\backslash T^{(\kappa)}$ and $T^{(\kappa)}$, for $\kappa=1,\ldots,K$. Each time, for fixed $\lambda$ and $\kappa$, we find the minimizer $\what{\bmgamma}^{(-\kappa)}(\lambda)$ of $\mathcal{L}(\bmgamma)+\sum_{j=1}^p p_\lambda(|\gamma_j|)$ using the training set $T\backslash T^{(\kappa)}$. Finally, we choose  $\lambda$ to be the minimizer of the following cross validation function
\[
\mbox{CV}(\lambda) = \sum_{\kappa=1}^K \mathcal{L}^{(\kappa)}(\what{\bmgamma}^{(-\kappa)}(\lambda)),
\]
where $\mathcal{L}^{(\kappa)}(\cdot)$ represents the evaluation of $\mathcal{L}(\cdot)$ using the test set $T^{(\kappa)}$.

Alternatively one can select $\lambda$ by the information criterion, for example, the generalized information criterion for high-dimensional penalized likelihood proposed by \cite{fan2013tuning}. They showed that the criterion with a uniform choice of the model complexity penalty identifies the true model with probability tending to 1 when the dimensionality grows at most exponentially fast with the sample size.

Although cross validation is computationally more expensive, it is less parsimonious and can often yield more satisfactory performance in practice. In this paper, we select the tuning parameter $\lambda$ by the $K$-fold cross validation with $K=5$.

\section{Theoretical Results}

We present the theoretical conditions and properties of our method for variable selection in the presence of missing data. For interpretation simplicity, we only show the results for a family of nonconvex penalties, including the SCAD and the MCP. The parallel results for the LASSO can be similarly developed and hence skipped. In fact, the assumptions for the LASSO to be selection consistent will be stronger \cite{zhao2006model}. The results we present hold under the high dimensional setting, that is the number of covariates can grow at most exponentially fast with the sample size.

\subsection{Notations}

The following notations are adopted throughout this paper.
For positive sequences $a_n$ and $b_n$, we denote $a_n \lesssim b_n$, if $a_n/b_n = O(1)$. We write $a_n \asymp b_n$ if $a_n \lesssim b_n$ and $b_n \lesssim a_n$.
For a vector $\bmv = (v_1, \ldots, v_p)^T \in R^p$, we define $\mbox{supp}(\bmv)=\{i: v_i\neq 0\}$, $|\mbox{supp}(\bmv)| = \mbox{card}\{\mbox{supp}(\bmv)\} = \|\bmv\|_0$, and $|A|$ is the cardinality of a set $A$.
For $1\leq q<\infty$, we define the $L_q$-norm as $\|\bmv\|_q = (\sum_{i=1}^p |v_i|^q)^{1/q}$.
Let $\|\bmv\|_\infty = \max_{1\leq i\leq p} |v_i|$ be the $L_\infty$-norm and $\bmv^{\otimes 2}=\bmv\bmv^T$ be the Kronecker product.
For two vectors $\bmv, \bmu \in R^p$, we denote $\bmv \circ \bmu = (v_1 u_1, \ldots, v_p u_p)^T$ as the Hadamard product.
For an $n\times p$ matrix $\matrM$, we define its matrix $L_1$-norm as $\|\matrM\|_{L_1} = \max_{1\leq j\leq p}\sum_{i=1}^n |M_{ij}|$, spectral norm as $\|\matrM\|_{2} = \sqrt{\lambda_{\rm max}(\matrM^T \matrM)}$, matrix $L_\infty$-norm as $\|\matrM\|_{L_\infty} = \max_{1\leq i\leq n}\sum_{j=1}^p |M_{ij}|$,
elementwise $L_1$-norm as $\|\matrM\|_1 = \sum_{i=1}^n \sum_{j=1}^p |M_{ij}|$, and elementwise supreme norm as $\|\matrM\|_{\infty} = \max_{i,j} \{|M_{ij}|\}$.
If $\matrM$ is squared and symmetric, we let $\lambda_{\rm min}(\matrM)$ and $\lambda_{\rm max}(\matrM)$ denote the minimal and maximal eigenvalues of $\matrM$.

The pairwise pseudo likelihood in (\ref{plike}) can be written as
\begin{eqnarray*}
\mathcal{L}(\bmgamma) &=& \frac{2}{n(n-1)}\sum_{1\leq i<j\leq n}\log\{1+\exp(-\yij\bmxij^T\bmgamma)\}\\
&=& \frac{2}{n(n-1)}\sum_{1\leq i<j\leq n}\{\psi(\yij\bmxij^T\bmgamma) - \yij\bmxij^T\bmgamma\},
\end{eqnarray*}
and its first and second order gradients are
\[
\nabla\mathcal{L}(\bmgamma) = \frac{2}{n(n-1)}\sum_{1\leq i<j\leq n}\{\psi'(\yij\bmxij^T\bmgamma
)\yij\bmxij - \yij\bmxij\},
\]
and
\[
\nabla^2\mathcal{L}(\bmgamma) = \frac{2}{n(n-1)}\sum_{1\leq i<j\leq n}\{\psi''(\yij\bmxij^T\bmgamma)\yij^2\bmxij^{\otimes 2}\},
\]
where we have $\psi(t)=\log(1+e^t)$, and hence $\psi'(t)=\frac{e^t}{1+e^t}$, $\psi''(t)=\frac{e^t}{(1+e^t)^2}$, $\psi'''(t)=\frac{e^t(1-e^t)}{(1+e^t)^3}$. After some algebra, it can be verified that the derivative functions are bounded. Particularly, we have $|\psi''(t)|\leq 0.25$ and $|\psi'''(t)|\leq 0.1$.

Throughout the paper, we denote the penalty function as $\mathcal{P}_\lambda(\bmgamma) = \sum_{j=1}^p p_\lambda(|\gamma_j|)$. We also define
$q_\lambda(t) = p_\lambda(t) - \lambda|t|$, $\mathcal{Q}_\lambda(\bmgamma) = \mathcal{P}_\lambda(\bmgamma) - \lambda\|\bmgamma\|_1$, and $\mathcal{\wtilde L}_\lambda(\bmgamma) = \mathcal{L}(\bmgamma) + \mathcal{Q}_\lambda(\bmgamma) = \mathcal{L}(\bmgamma) + \mathcal{P}_\lambda(\bmgamma) - \lambda\|\bmgamma\|_1$. Therefore, the penalized objective function in (\ref{plike}) can be written as
\begin{eqnarray}
\mathcal{L}(\bmgamma) + \mathcal{P}_\lambda(\bmgamma) = \mathcal{\wtilde L}_\lambda(\bmgamma) + \lambda \|\bmgamma\|_1.
\label{plike2}
\end{eqnarray}

We denote $\bmtheta^\ast$ as the true value of parameter $\bmtheta$ and
$\bmgamma^\ast = (\gamma_1^\ast, \ldots, \gamma_p^\ast)^T$ as the true value of $\bmgamma$.
We define $S=\{j: \gamma_j^\ast \neq 0\} = \{j: \beta_j^\ast \neq 0\}$, and its complement $\bar{S}=\{j: \gamma_j^\ast = 0\} = \{j: \beta_j^\ast = 0\}$, where $s^\ast=|S|<n$.
For any vector $\bmxi \in R^p$, we define $\bmxi_S = \{v_j: j \in S\} \in R^{s^\ast}$.
For the $p \times p$ Hessian matrix $\nabla^2\mathcal{L}(\bmgamma)$, we write $\nabla_{SS}^2\mathcal{L}(\bmgamma)$ as the corresponding $s^\ast \times s^\ast$ sub-matrix with restrictions to the coordinates in $S$.
Finally, the oracle estimator is defined as
\[
\what{\bmgamma}_{\rm O} = \argmin_{\mbox{supp}(\bmgamma)\subset S, \bmgamma\in R^{p}} \mathcal{L}(\bmgamma).
\]

\subsection{Assumptions}

In this subsection, we present the main assumptions that are necessary to derive our theoretical results. Our first assumption is on how to control the tail behavior of $Y$ given $\bmX$.
\begin{assumption}
  Assume that $\|\bmX\|_{\infty}<M<\infty$, $|\bmX^T\bmgamma^\ast|<B<\infty$, and $Y$ given $\bmX$ has the sub-exponential tail, i.e., for any $\delta>0$, $\Pr(|Y|\geq \delta|\bmX)\leq c_1\exp(-c_2\delta)$, where $c_1$ and $c_2$ are positive constants.
\end{assumption}
This assumption is similar to the Assumption 3.7 in \cite{ning2017likelihood}. As they verified,
the sub-exponential tail assumption is satisfied for most commonly used GLMs in practice, for example, linear regression with Gaussian noise and logistic regression.

Next, we need to impose some conditions on the extreme sparse eigenvalues of a matrix $\matrM$. We first define sparse eigenvalues as follows.
\begin{definition}
  Let $s$ be a positive integer. The largest and smallest $s$-sparse eigenvalues of a $p$-dimensional squared matrix $\matrM$ are
\[
\rho_+(\matrM, s) = \sup\{\bmv^T \matrM \bmv: \|\bmv\|_0\leq s, \|\bmv\|_2=1\},
\]
and
\[
\rho_-(\matrM, s) = \inf\{\bmv^T \matrM \bmv: \|\bmv\|_0\leq s, \|\bmv\|_2=1\}.
\]
\end{definition}
Since we frequently use $\rho_+(\nabla^2\mathcal{L}(\bmgamma^\ast), s)$ and $\rho_-(\nabla^2\mathcal{L}(\bmgamma^\ast), s)$ in the following derivation, we simply rewrite them as $\rho_+(s) = \rho_+(\nabla^2\mathcal{L}(\bmgamma^\ast), s)$ and $\rho_-(s) = \rho_-(\nabla^2\mathcal{L}(\bmgamma^\ast), s) $.
\begin{assumption}
  There exists positive constants $\rho_\ast$ and $\rho^\ast$, such that
  \[
  \rho_\ast \leq \rho_-(s) \leq \rho_+(s) \leq \rho^\ast.
  \]
\end{assumption}
The sparse eigenvalue conditions are usually proposed to bound the estimation error in high dimensional problems. The similar concepts, although in slightly different forms, have been defined and studied in \cite{bickel2009simultaneous, ning2017likelihood, yang2014semiparametric}. In Appendix, we verify that the Assumption 2 is satisfied with probability at least $1-C_1 p^2 \exp(-C_2 n/s^2)$ for most commonly used GLMs, i.e., linear regression with Gaussian noise and logistic regression.
The definitions of $C_1$ and $C_2$ are in the Appendix.


Recall that the theory presented in this Section applies not only to the SCAD and the MCP penalties, instead, it applies to a broad  class of nonconvex penalties. We rely on the following regularity conditions for penalty functions.
\begin{assumption}
  Assume the following conditions are satisfied for $p_\lambda(t)$ or $q_\lambda(t)$ or their first order derivatives:
  \begin{itemize}
    \item[(a)] $q_\lambda(t)$ is symmetric, i.e., $q_\lambda(-t) = q_\lambda(t)$ for any $t$, and $q_\lambda(0) = 0$;
    \item[(b)] $q'_\lambda(t)$ is monotone and Lipschitz continuous, i.e., for $t'>t$, there exist two constants $\zeta_- \geq 0$ and $\zeta_+ \geq 0$ such that
        \[
        -\zeta_- \leq \frac{q'_\lambda(t') - q'_\lambda(t)}{t'-t} \leq -\zeta_+ \leq 0;
        \]
    \item[(c)] $q'_\lambda(t)$ is bounded, i.e., $|q'_\lambda(t)|\leq \lambda$ for any $t$, and $q'_\lambda(0) = 0$;
    \item[(d)] $q'_\lambda(t)$ has bounded difference with respect to $\lambda$: $|q'_{\lambda_1}(t) - q'_{\lambda_2}(t)| \leq |\lambda_1 - \lambda_2|$ for any $t$;
    \item[(e)] There exist $c_7 \in [0,1]$ and $c_8 \in (0,\infty)$ such that $q'_\lambda(t) \geq (c_7-1)\lambda$, i.e., $p'_\lambda(t) \geq c_7 \lambda$ for $t \in (0,c_8 \lambda]$;
    \item[(f)] $p'_\lambda(t)=0$ once $|t|>\nu>c_9\sqrt{\frac{\log p}{n}}$ for some positive constant $c_9$.
  \end{itemize}
\end{assumption}
The assumptions presented here are similar to \cite{wang2014optimal, yang2014semiparametric}. In $(b)$, $\zeta_-$ and $\zeta_+$ are two parameters that control the concavity of $q_\lambda(t)$. Taking $t'\rightarrow t$ in $(b)$, we have $q''_\lambda(t)\in [-\zeta_-, -\zeta_+]$, which suggests that larger $\zeta_-$ and $\zeta_+$ allow $q_\lambda(t)$ to be more concave. For example, in SCAD we have $\zeta_- = 1/(a-1)$ with some $a>2$ and $\zeta_+=0$, and in MCP we have $\zeta_- = 1/a$ with some $a>0$ and $\zeta_+=0$. In \cite{wang2014optimal}, they illustrated that all the conditions hold for both SCAD and MCP.

In some of our following derivations, we also need a relation between the concavity parameter $\zeta_-$ and $\rho_-(\nabla^2\mathcal{L}, 2s^\ast)$, the smallest $(2s^\ast)$-sparse eigenvalue of the Hessian matrix $\nabla^2\mathcal{L}$.
\begin{assumption}
  The concavity parameter $\zeta_-$ defined in the conditions for the penalty function satisfies
  \[
  \zeta_-  \leq c_{10}\rho_-(\nabla^2\mathcal{L}, 2s^\ast),
  \]
  with some constant $c_{10}<1$.
\end{assumption}
Since in fact $\zeta_+ \leq \zeta_-$ and $\rho_-(\nabla^2\mathcal{L}, 2s^\ast) \leq \rho_+(\nabla^2\mathcal{L}, 2s^\ast)$, the restriction above implies that $\zeta_+  \leq c_{10}\rho_+(\nabla^2\mathcal{L}, 2s^\ast)$. Theoretically, for each penalty, these two restrictions are satisfied by going through the verification of the Assumption 2 and appropriately choose the $t$ value and $\rho_\ast$, $\rho^\ast$ values.

\subsection{Main Results}

Our main objective in this subsection is to show that, the estimator from our proposed method, $\what\bmgamma$, has the same support as the true value $\bmgamma^\ast$, also as $\bmbeta^\ast$, i.e., the variable selection consistency property satisfies. A sequence of results will be presented in the following.
The first result shows that the true value of $\bmgamma$, $\bmgamma^\ast$, minimizes $\E(\mathcal{L}(\bmgamma))$. Recall that $\mathcal{L}(\bmgamma)$ is the unpenalized component of the objective function, defined in $(\ref{acloglike})$. This result provides the intuition why $\mathcal{L}(\bmgamma)$ is a legitimate loss function.
\begin{lemma}
\label{globalminimizer}
  We have $\E(\nabla \mathcal{L}(\bmgamma^\ast))=0$ and $\bmgamma^\ast$ is a global minimizer of $\E(\mathcal{L}(\bmgamma))$, where $\E(\cdot)$ is the expectation under the true parameter $\bmtheta^\ast$.
\end{lemma}

Recall that $\nabla\mathcal{L}(\bmgamma)$ has a second-order U-statistic structure.
Our second result concerns the concentration inequality for U-statistics with a sub-exponential kernel function. We only present the result for second-order U-statistics. In \cite{ning2017likelihood, yang2014semiparametric}, the authors provided a more general concentration inequality.
\begin{lemma}
\label{concentration}
  Let $X_1, \ldots, X_n$ be independent random variables. Consider the following U-statistics of order 2
  \[
  U_n = \frac{2}{n(n-1)}\sum_{1\leq i_1<i_2 \leq n} u(X_{i_1}, X_{i_2}),
  \]
  where $\E\{u(X_{i_1}, X_{i_2})\}=0$ for all $i_1<i_2$. If there exist constants $L_1$ and $L_2$ such that
  \[
  \Pr(|u(X_{i_1}, X_{i_2})|\geq x) \leq L_1 \exp(-L_2 x),
  \]
  for all $i_1<i_2$ and all $x\geq 0$, then
  \[
  \Pr(|U_n|\geq x) \leq 2\exp\left[ -\min\left\{ \frac{L_2^2 x^2}{8L_1^2}, \frac{L_2 x}{4L_1} \right\}k \right],
  \]
  where $k=\lfloor n/2 \rfloor$ is the largest integer less than $n/2$.
\end{lemma}

The proofs of the above two results are contained in previous literature, see \cite{ning2017likelihood, yang2014semiparametric}, so they are omitted. The next result controls the magnitude of $\|\nabla\mathcal{L}(\bmgamma^\ast)\|_\infty$.

\begin{lemma}
\label{magnitude}
  Given the Assumption 1, we have
  \[
  \|\nabla\mathcal{L}(\bmgamma^\ast)\|_\infty \leq C_3 \sqrt{\log p/n},
  \]
  with probability at least $1 - \delta_1$, where $\delta_1 = 2p \exp\left[ -\min \left\{ C_4\log p, C_5 n^{1/2} (\log p)^{1/2} \right\} \right]$, where $C_3$ is a positive constant, $C_4$ and $C_5$ are constants defined in detail in the Appendix.
\end{lemma}


Based on the magnitude of $\|\nabla\mathcal{L}(\bmgamma^\ast)\|_\infty$, we can provide a bound for the difference between the truth $\bmgamma^\ast$ and the oracle estimator $\what{\bmgamma}_{\rm O}$, as follows.

\begin{lemma}
\label{oraclediff}
Given the Assumption 1 and the assumption that
$\|\nabla_{SS}^2\mathcal{L}(\bmgamma^\ast)^{-1}\|_{L_\infty} < C$,
$\log(n)(s^\ast)^2 \sqrt{\frac{\log p}{n}}=o(1)$,
we have
  \[
\|\what{\bmgamma}_{\rm O} - \bmgamma^\ast\|_\infty < 2C C_3 \sqrt{\frac{\log s^\ast}{n}},
\]
with probability at least $1- \delta_2$, where $\delta_2 = 2s^{\ast} \exp\left[ -\min \left\{ C_4\log s^{\ast}, C_5 n^{1/2} (\log s^{\ast})^{1/2} \right\} \right] +c_{12} p^{-1} +c_1 n^{-1}$.
\end{lemma}

Next, we present a characteristic of our surrogate loss function that satisfies the restricted strong convexity and restricted strong smoothness properties when evaluated at two sparse vectors $\bmgamma_1$ and $\bmgamma_2$ satisfying  $\|(\bmgamma_1-\bmgamma_2)_{\bar{S}}\|_0\leq s^\ast$, i.e., for the coordinates in $\bar{S}$, the cardinality of the support set of $\bmgamma_1-\bmgamma_2$ is bounded by the true number of ``important'' variables, which actually bounds the false positive magnitude.

\begin{lemma}
\label{restricted}
Given the Assumption 3, if $\bmgamma_1$ and $\bmgamma_2$ are two $p$-dimensional sparse vectors, which satisfy $\|(\bmgamma_1-\bmgamma_2)_{\bar{S}}\|_0\leq s^\ast$, then the surrogate loss function satisfies the restricted strong convexity
\[
\mathcal{\wtilde L}_\lambda(\bmgamma_2) \geq \mathcal{\wtilde L}_\lambda(\bmgamma_1) + \nabla \mathcal{\wtilde L}_\lambda(\bmgamma_1)^T (\bmgamma_2-\bmgamma_1) + \frac{\rho_-(\nabla^2\mathcal{L}, 2s^\ast) - \zeta_-}{2}\|\bmgamma_2-\bmgamma_1\|_2^2,
\]
and the restricted strong smoothness
\[
\mathcal{\wtilde L}_\lambda(\bmgamma_2) \leq \mathcal{\wtilde L}_\lambda(\bmgamma_1) + \nabla \mathcal{\wtilde L}_\lambda(\bmgamma_1)^T (\bmgamma_2-\bmgamma_1) + \frac{\rho_+(\nabla^2\mathcal{L}, 2s^\ast) - \zeta_+}{2}\|\bmgamma_2-\bmgamma_1\|_2^2.
\]
\end{lemma}

Finally, we present our variable selection consistency result. We achieve this goal by showing the support set of our proposed estimator and that of the oracle estimator are the same as that of the true parameter.

\begin{theorem}
\label{major}
Assume that the Assumptions 1, 2, 3, 4 hold, $\|\nabla_{SS}^2\mathcal{L}(\bmgamma^\ast)^{-1}\|_{L_\infty} < C$, where $C$ is a positive constant specified in Lemma 4, $\log(n)(s^\ast)^2 \sqrt{\frac{\log p}{n}}=o(1)$, and the weakest signal strength satisfies $\min_{j \in S}|\gamma_j^\ast|>2\nu>2\lambda$, where $\lambda \asymp \sqrt{\log p/n}$. Then, when $n$ is sufficiently large, we have $\what\bmgamma=\what\bmgamma_{\rm O}$, and hence
\[
\mbox{supp}(\what\bmgamma)=\mbox{supp}(\what\bmgamma_{\rm O})=\mbox{supp}(\bmgamma^\ast),
\]
with probability at least $1 - \delta_1 - \delta_2 - \delta_3$,
where $\delta_1$ is defined in Lemma 3, $\delta_2$ is defined in Lemma 4, $\delta_3 = C_1 p^2 \exp(-C_2 n/(s^\ast)^2)$ comes from the Assumption 2.
\end{theorem}

\begin{remark}
  In Theorem 1, the lower bound of the high probability comes from those in Assumption 2, Lemma 3 and Lemma 4. To be more specific, using Lemma 4, in equation (1) in the proof, we show that $|(\what\bmgamma_{\rm O})_j|>\nu$ with probability at least $1- \delta_2$; using Lemma 3, in equation (2) in the proof, we have $\|\nabla\mathcal{L}(\what\bmgamma_{\rm O})\|_\infty \leq C_3 \sqrt{\log p/n}$ with probability at least $1 - \delta_1$; based on the Assumption 2, we establish $\|\what\bmgamma^{(l)}-\bmgamma^\ast\|_2 \leq c_{14}\rho_\ast^{-1}\sqrt{s^\ast}\lambda$ in equation (3) with probability at least $1-\delta_3$. The final lower bound of the high probability comes from the combination of all three together and the fact that $P(A\bigcap B\bigcap C)\geq P(A) +P(B\bigcap C)-1 \geq P(A) +P(B) + P(C)-2 \geq 1-\delta_1 - \delta_2 - \delta_3$ where $A$, $B$ and $C$ are three arbitrary events, and $P(A)\geq 1-\delta_1$, $P(B)\geq 1-\delta_2$, $P(C)\geq 1-\delta_3$.
\end{remark}

\begin{remark}
  With respect to the high dimensional set-up, we allow both $\log p$, the logarithm of the dimensionality, and $s^\ast$, the number of nonzero components in the original parameter $\bfbeta$, to grow with $n$. From Theorem 1 and its proof, the condition $\log p$ and $s^\ast$ need to be satisfied is that $\log(n)(s^\ast)^2 \sqrt{\frac{\log p}{n}}=o(1)$.
  It implies that if $s^\ast = o(n^\varsigma)$ for some $0<\varsigma<1/4$, then $\log p = o(n^{1-4\varsigma}/(\log n)^2)$.
  Note that we follow the most recent statistical literature for the definition of high dimensionality. For example, in \citep{fan2011nonconcave}, the high dimensionality refers to $\log p = O(n^{\alpha})$, for some $0<\alpha<1$. Here we have $\log p = o(n^{1-4\varsigma}/(\log n)^2)$. It means, in the high dimensional GLM as we consider, the number of covariates $p$ can grow at most exponentially fast with $n$, the sample size of the completely observed subjects.
\end{remark}

\section{Simulation Studies}

The objective of our simulation studies is two-fold. First, we evaluate the finite sample performance of our proposed method by examining two commonly used models: linear regression and logistic regression, and three representative penalty functions: LASSO, SCAD and MCP. Second, we compare our proposed method to two existing methods: one assuming that there is no missing data, and the other assuming the missing data mechanism is MAR.

In all of our six simulation settings (S1)--(S6), we generate the covariate $\bmX$ from $p$-dimensional $N(\bmzero, \bmSigma)$, where $\bmSigma_{ij} = \rho^{|i-j|}$, and we consider $\rho \in \{0, 0.5\}$. Recall the function $b(\eta)$ in the definition (\ref{glm}) of GLM: $b(\eta)=\eta^2/2$ corresponds to linear regression and $b(\eta)=\log(1+e^\eta)$ corresponds to logistic regression.

Our simulation settings (S1)--(S4) are as follows:\\
(S1): $b(\eta)=\eta^2/2$, with $\eta=\alpha+\bmbeta^T \bmX$, $\alpha=0$, $\bmbeta=(3, 1.5, 0.5, 0, \ldots, 0)^T$, the dispersion parameter $\phi=1$, $s^\ast=3$, $p=8$ and $N=200$. The missing data mechanism $\Pr(R=1|Y, \bmX) = I_{\{Y>\gamma_1\}} I_{\{X_1>\gamma_2\}}$ with $\gamma_1 = -3.3$, $\gamma_2 = -0.4$ for $\rho=0$ and $\gamma_1 = -3.8$, $\gamma_2 = -0.3$ for $\rho=0.5$.\\
(S2): $b(\eta)=\eta^2/2$, with $\eta=\alpha+\bmbeta^T \bmX$, $\alpha=0$, $\bmbeta=(3, 1.5, 0.5, 0, \ldots, 0)^T$, the dispersion parameter $\phi=1$, $s^\ast=3$, $p=200$ and $N=200$. The missing data mechanism $\Pr(R=1|Y, \bmX) = I_{\{Y>\gamma_1\}} I_{\{X_1>\gamma_2\}}$ with $\gamma_1 = -2.8$, $\gamma_2 = -0.4$ for $\rho=0$ and $\gamma_1 = -4.1$, $\gamma_2 = -0.3$ for $\rho=0.5$.\\
(S3): $b(\eta)=\log(1+e^\eta)$, with $\eta=\alpha+\bmbeta^T \bmX$, $\alpha=0$, $\bmbeta=(2, -2, 1, -1, 0, \ldots, 0)^T$, $s^\ast=4$, $p=8$ and $N=500$. The missing data mechanism $\Pr(R=1|Y, \bmX) = I_{\{X_1>\gamma\}}\cdot(2Y+3)/5$ with $\gamma = -0.7$ for either $\rho=0$ or $\rho=0.5$.\\
(S4): $b(\eta)=\log(1+e^\eta)$, with $\eta=\alpha+\bmbeta^T \bmX$, $\alpha=0$, $\bmbeta=(2, -2, 1, -1, 0, \ldots, 0)^T$, $s^\ast=4$, $p=500$ and $N=500$. The missing data mechanism $\Pr(R=1|Y, \bmX) = I_{\{X_1>\gamma\}}\cdot(2Y+3)/5$ with $\gamma = -0.7$ for either $\rho=0$ or $\rho=0.5$.

\begin{figure}[htbp]
\centering
\includegraphics[scale=.25]{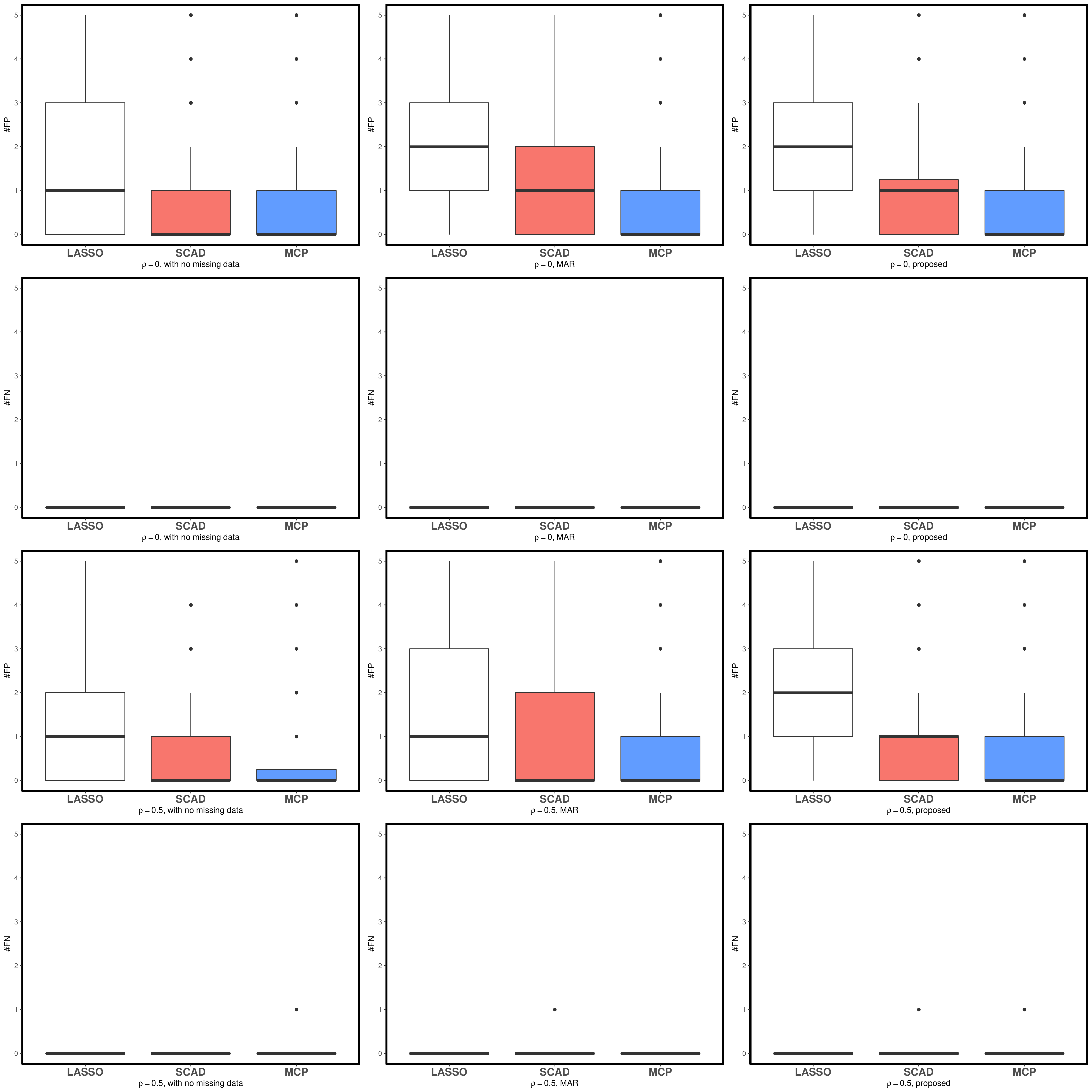}
\caption{Boxplots of $\#$FP and $\#$FN in simulation setting (S1). The three columns represent the methods with no missing data, MAR and proposed, respectively. The first and third rows show $\#$FP while the second and fourth rows show $\#$FN. The first two rows are for the case with $\rho=0$ and the last two rows are for the case with $\rho=0.5$.}
\label{figure:linearlow}
\end{figure}

\begin{figure}[htbp]
\centering
\includegraphics[scale=.25]{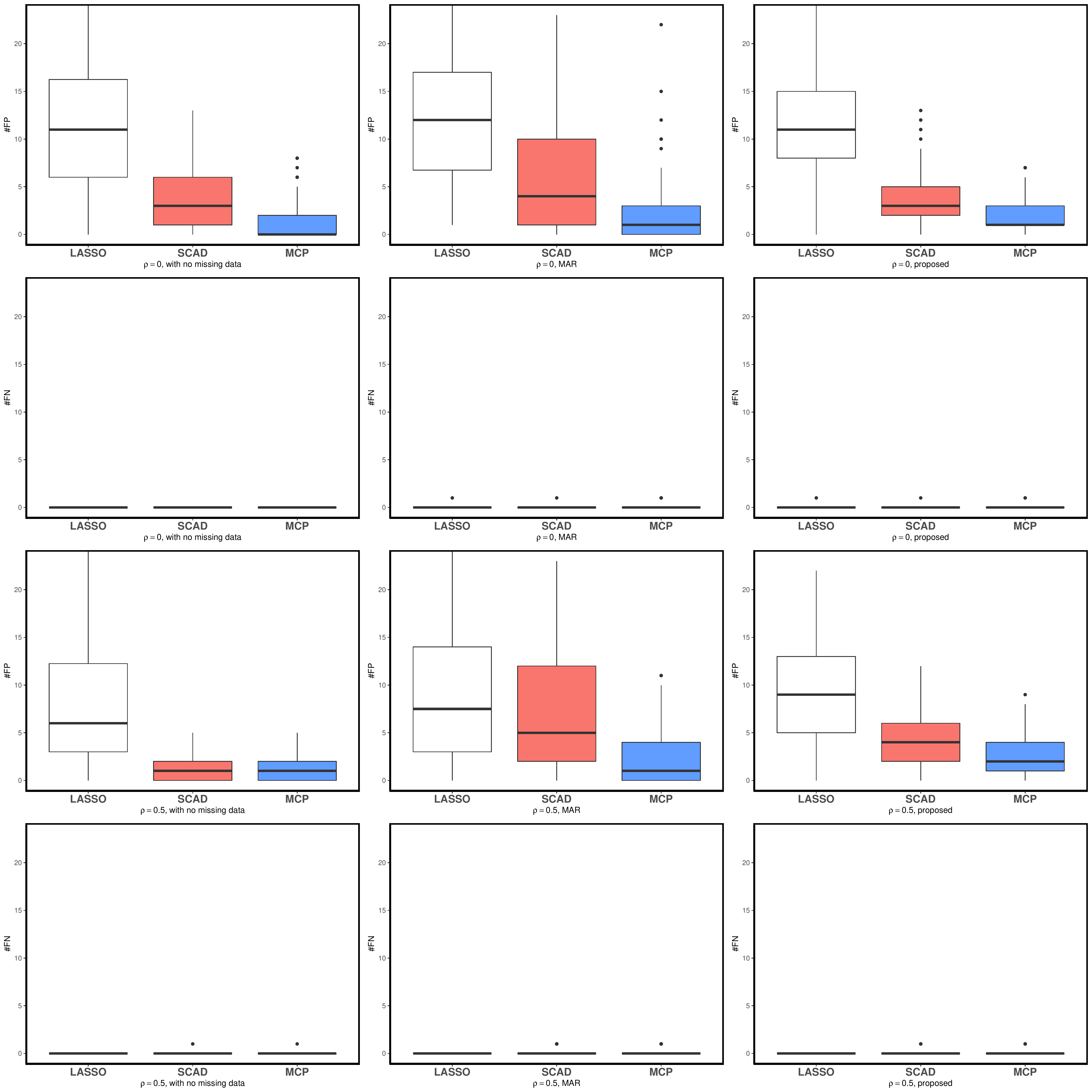}
\caption{Boxplots of $\#$FP and $\#$FN in simulation setting (S2). The three columns represent the methods with no missing data, MAR and proposed, respectively. The first and third rows show $\#$FP while the second and fourth rows show $\#$FN. The first two rows are for the case with $\rho=0$ and the last two rows are for the case with $\rho=0.5$.}
\label{figure:linearhigh}
\end{figure}

\begin{figure}[htbp]
\centering
\includegraphics[scale=.25]{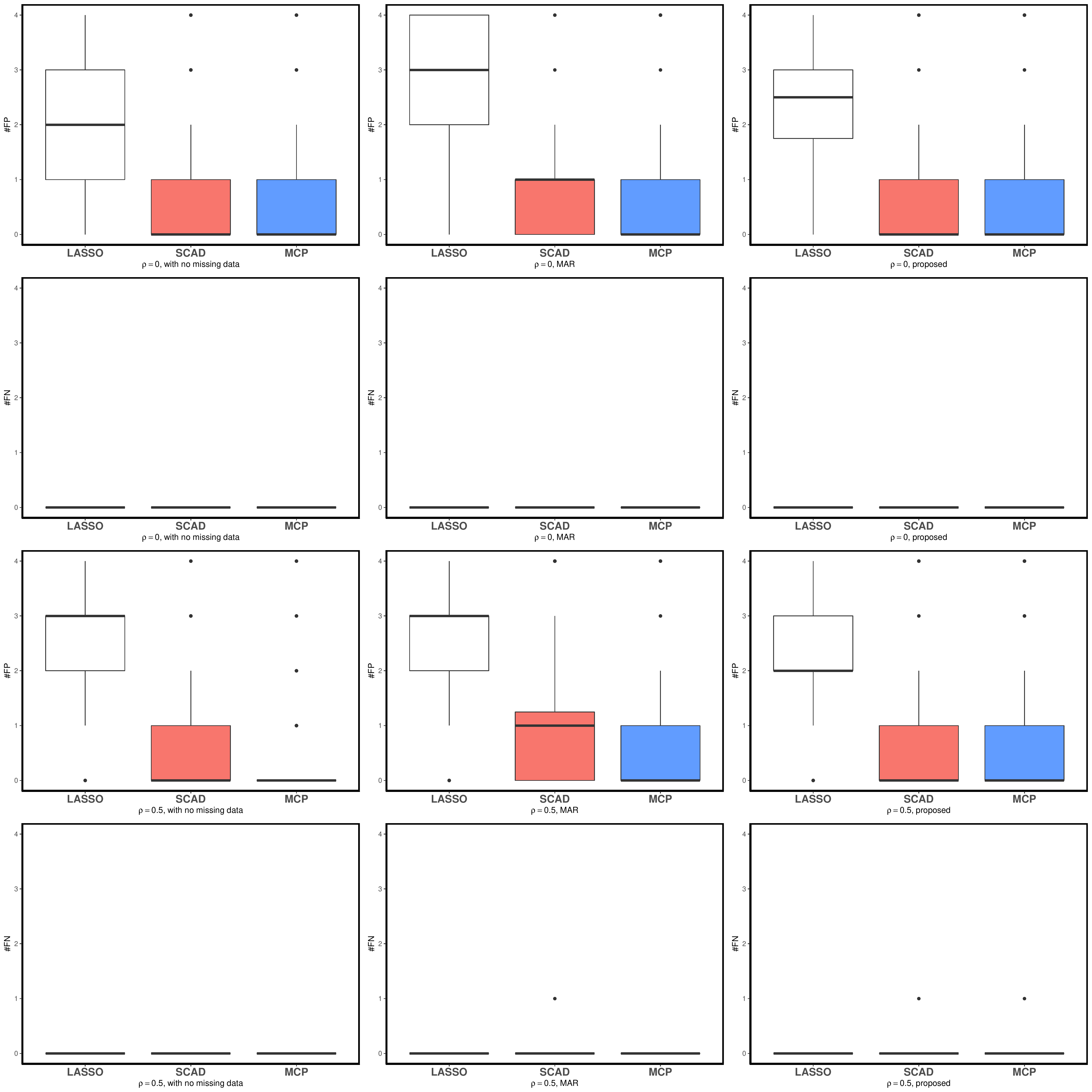}
\caption{Boxplots of $\#$FP and $\#$FN in simulation setting (S3). The three columns represent the methods with no missing data, MAR and proposed, respectively. The first and third rows show $\#$FP while the second and fourth rows show $\#$FN. The first two rows are for the case with $\rho=0$ and the last two rows are for the case with $\rho=0.5$.}
\label{figure:logisticlow}
\end{figure}

\begin{figure}[htbp]
\centering
\includegraphics[scale=.25]{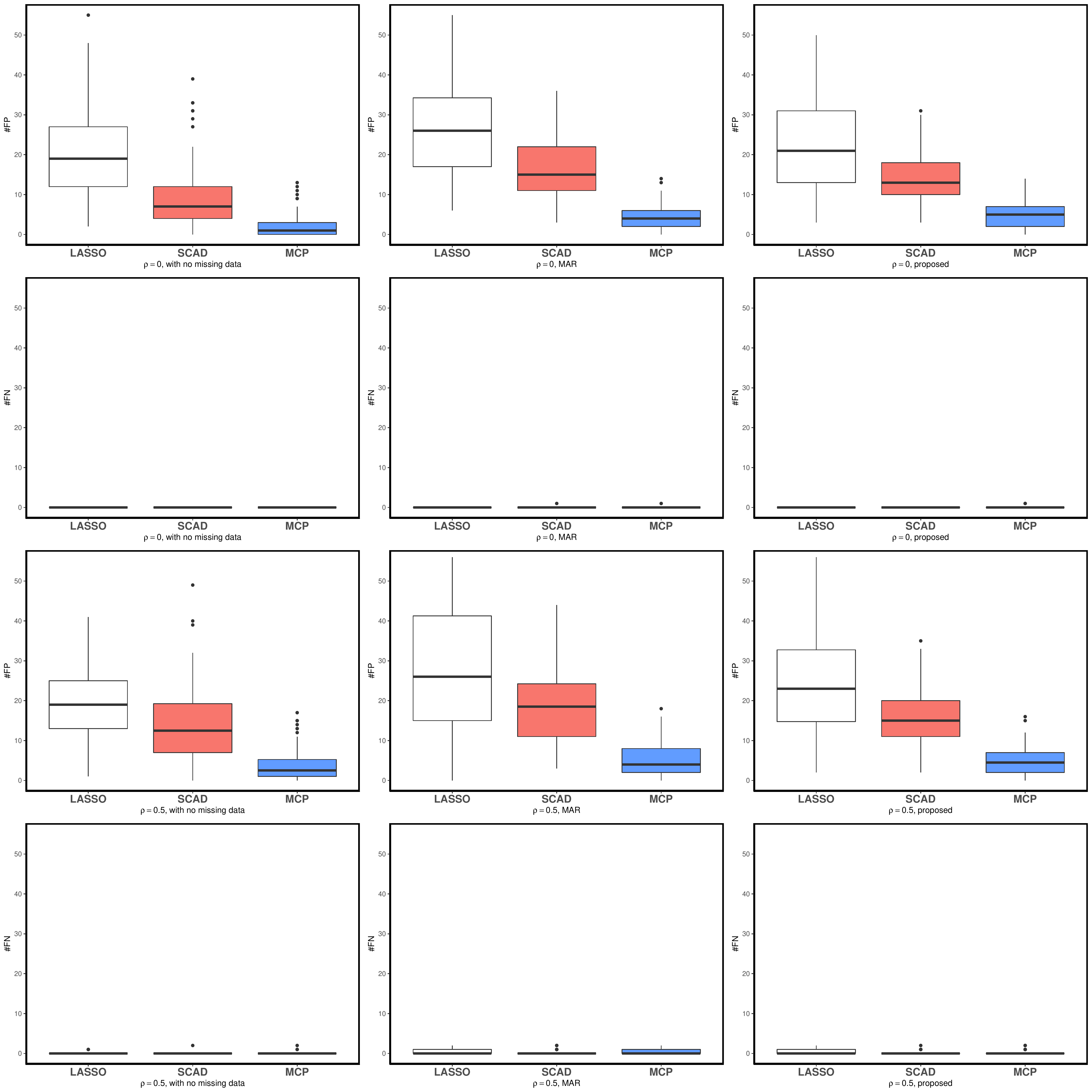}
\caption{Boxplots of $\#$FP and $\#$FN in simulation setting (S4). The three columns represent the methods with no missing data, MAR and proposed, respectively. The first and third rows show $\#$FP while the second and fourth rows show $\#$FN. The first two rows are for the case with $\rho=0$ and the last two rows are for the case with $\rho=0.5$.}
\label{figure:logistichigh}
\end{figure}

The purpose of the different choices of $\gamma$ values is to guarantee that, in each setting, the observed proportion is about 60$\%$ to 65$\%$.
We report the results based on 100 replications in each setting. We define false positive (FP) as the one with true zero value but falsely estimated as nonzero; and false negative (FN) as the one with true nonzero value but falsely estimated as zero. We count the number of false positives ($\#$FP) and the number of false negatives ($\#$FN) and report them in a boxplot in each setting in Figures \ref{figure:linearlow}--\ref{figure:logistichigh} respectively. We also list the mean and standard deviation (SD) of $\#$FP and $\#$FN for each setting in Tables \ref{table:linear}--\ref{table:logistic} for linear regression and logistic regression, respectively.

Some conclusions can be reached from simulation studies (S1)--(S4).
First, in almost all scenarios, our proposed method outperforms the method assuming MAR in terms of smaller FP and FN mean/median values.
Second, in most scenarios, the method with no missing data, treated as a gold standard, outperforms our proposed method.
Third, the nonconvex penalties almost always perform better than the LASSO penalty in terms of variable selection, which is consistent with the previous literature.

Under the assumption (\ref{assume}), the method assuming MAR produces biased estimators and hence worse results for variable selection, while the proposed estimator satisfies the variable selection consistency property and hence better (than the MAR method) variable selection performance is expected. Therefore, our numerical findings in (S1)--(S4) well match the theory.

In general the assumption imposed on the missing data mechanism is unverifiable. Although the assumption (\ref{assume}) we discuss in this paper is already very flexible, it is still plausible to be violated in real applications. Therefore, in the next two simulations, we evaluate the robustness of our proposed method when the assumption (\ref{assume}) is slightly violated. The simulation settings (S5)--(S6) are as follows:\\
(S5): $b(\eta)=\eta^2/2$, with $\eta=\alpha+\bmbeta^T \bmX$, $\alpha=0$, $\bmbeta=(3, 1.5, 0.5, 0, \ldots, 0)^T$, the dispersion parameter $\phi=1$, $s^\ast=3$, $p=8$ and $N=200$. The missing data mechanism $\Pr(R=1|Y, \bmX) = I_{\{Y+0.1 X_3>\gamma_1\}} I_{\{X_1>\gamma_2\}}$ with $\gamma_1 = -3.3$, $\gamma_2 = -0.4$ for $\rho=0$ and $\gamma_1 = -3.8$, $\gamma_2 = -0.3$ for $\rho=0.5$.\\
(S6): $b(\eta)=\eta^2/2$, with $\eta=\alpha+\bmbeta^T \bmX$, $\alpha=0$, $\bmbeta=(3, 1.5, 0.5, 0, \ldots, 0)^T$, the dispersion parameter $\phi=1$, $s^\ast=3$, $p=200$ and $N=200$. The missing data mechanism $\Pr(R=1|Y, \bmX) = I_{\{Y+0.1 X_3>\gamma_1\}} I_{\{X_1>\gamma_2\}}$ with $\gamma_1 = -2.8$, $\gamma_2 = -0.4$ for $\rho=0$ and $\gamma_1 = -4.1$, $\gamma_2 = -0.3$ for $\rho=0.5$.

\begin{figure}[htbp]
\centering
\includegraphics[scale=.25]{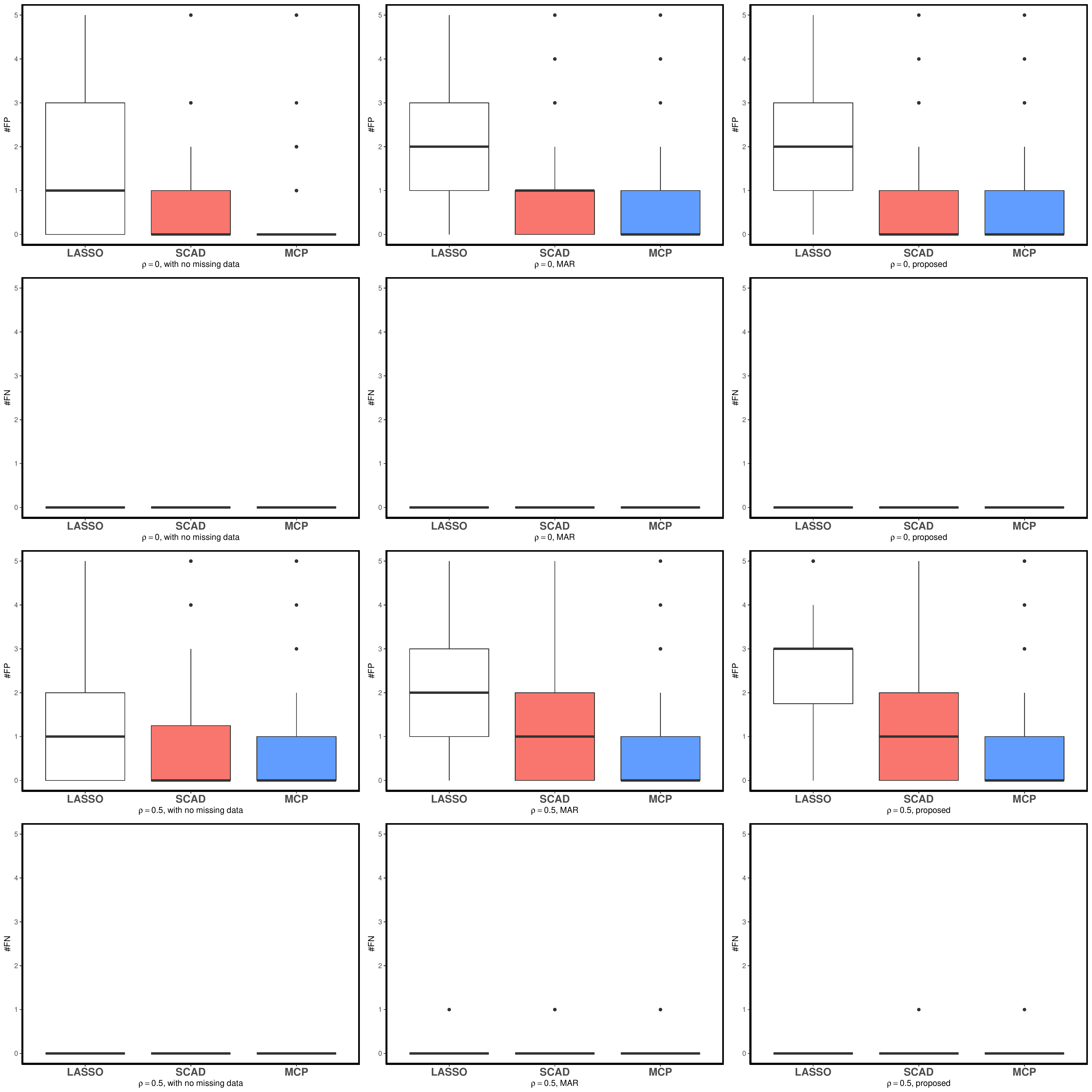}
\caption{Boxplots of $\#$FP and $\#$FN in simulation setting (S5). The three columns represent the methods with no missing data, MAR and proposed, respectively. The first and third rows show $\#$FP while the second and fourth rows show $\#$FN. The first two rows are for the case with $\rho=0$ and the last two rows are for the case with $\rho=0.5$.}
\label{figure:linearlow2}
\end{figure}

\begin{figure}[htbp]
\centering
\includegraphics[scale=.25]{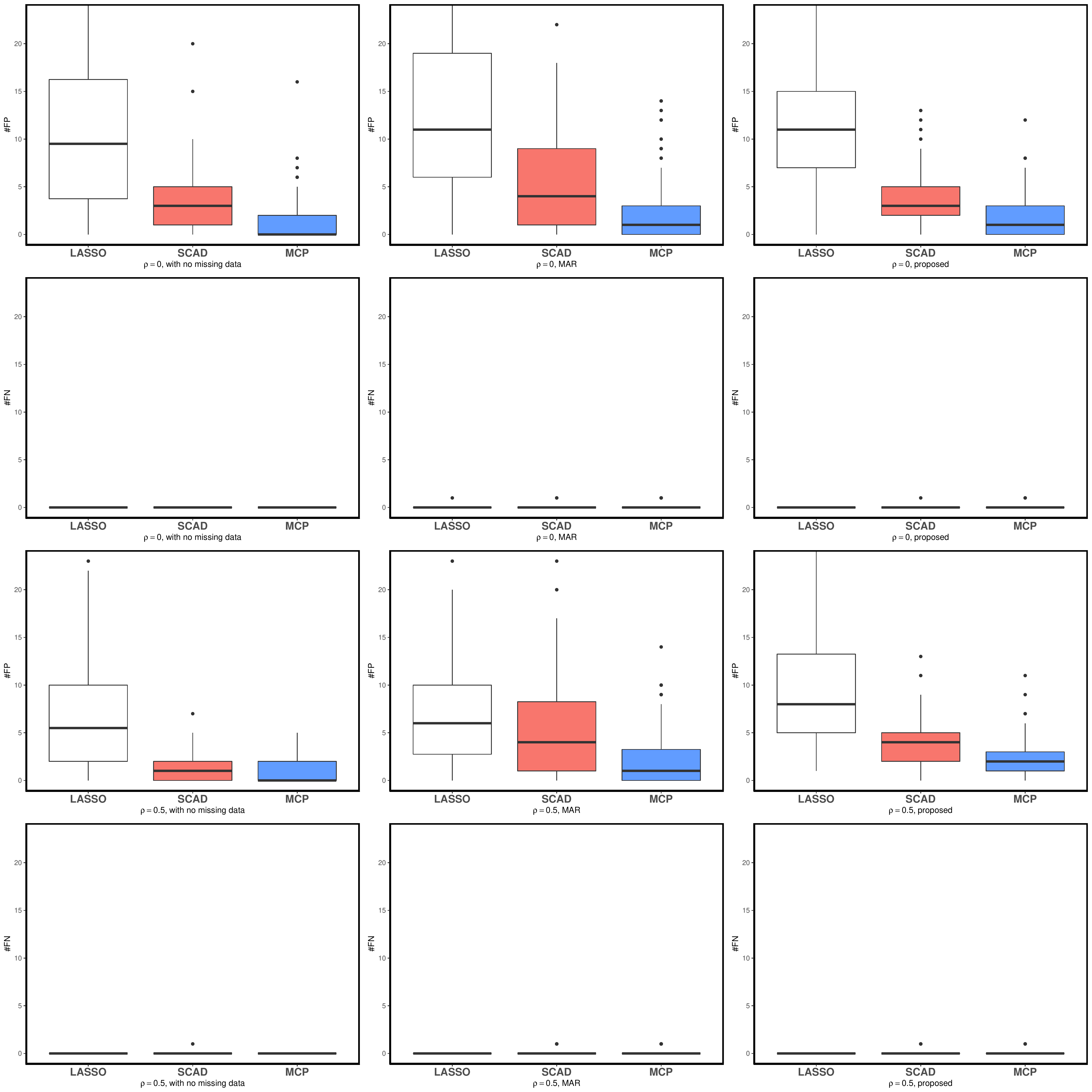}
\caption{Boxplots of $\#$FP and $\#$FN in simulation setting (S6). The three columns represent the methods with no missing data, MAR and proposed, respectively. The first and third rows show $\#$FP while the second and fourth rows show $\#$FN. The first two rows are for the case with $\rho=0$ and the last two rows are for the case with $\rho=0.5$.}
\label{figure:linearhigh2}
\end{figure}

Similar as before, we count the number of false positives ($\#$FP) and the number of false negatives ($\#$FN) and report them in a boxplot in each setting in Figures \ref{figure:linearlow2}--\ref{figure:linearhigh2} respectively. We also list the mean and standard deviation (SD) of $\#$FP and $\#$FN for each setting in Table \ref{table:linear2}.
It can be seen that, although the assumption (\ref{assume}) is slightly violated, our proposed method still performs better than the one assuming MAR in many scenarios. This phenomenon shows that our proposed method possesses some robustness to the misspecification of the missing data mechanism assumption.

Finally, we provide some results on the computing time of our proposed method.
We report the mean and standard deviation (SD) of the computing time for simulation settings (S1)--(S2) in Table \ref{table:time}. The simulations are conducted on an OS X system version 10.9.5 with 2.2 GHz Intel Core i7 CPU and 16GB memory.
It's not surprising that our proposed method is more time-consuming than the others.
This phenomenon is consistent with the theoretical implication.
In theory, from the algorithms we developed in Section 3, the computing time of the proposed method is equivalent to solving a standard penalized logistic regression with sample size $n(n-1)/2$, while the computing time of the method assuming no missing data (or assuming MAR) is the same as to solving a standard penalized logistic regression with sample size $N$ (or $n$).
Eventually we will make our algorithm publicly available by creating an R package with some core part implemented by C.

\section{Real Data Analyses}

In this Section, we present two data analyses to demonstrate the usefulness of our proposed method in real applications. The first study concerns the melanoma cancer through the observation-controlled Eastern Cooperative Oncology Group (ECOG) phase III clinical trial E1684. The second study (GEO GDS3289) investigates the association between prostate cancer tumors and genomic biomarkers, sponsored by the US National Institutes of Health.

\subsection{Melanoma Study}

Melanoma is the most dangerous type of skin cancer and its incidence is increasing at a rate that exceeds all solid tumors.
Although education efforts have resulted in earlier detection of melanoma, high-risk melanoma patients continue to have high relapse and mortality rate of 50$\%$ or higher.
Several post-operative (adjuvant) chemotherapies have been proposed for this class of melanoma patients, and the one which seems to provide the most significant impact on relapse-free survival and survival is Interferon Alpha-2b (IFN). This immunotherapy was evaluated in E1684, an observation-controlled Eastern Cooperative Oncology Group (ECOG) phase III clinical trial \cite{kirkwood1996interferon}.

In this trial, there are in total $N=286$ patients and all the patients were randomized to one of two treatment trials: high dose interferon or observation. In this analysis, the outcome variable $Y$, was taken to be binary, and was assigned a 1 if the patient had an overall survival time greater than or equal to 0.55 years, and 0 otherwise. There are several prognostic factors that were identified as potentially important predictors: $X_1$, treatment (two levels); $X_2$, age (in years); $X_3$, nodes1 (four levels); $X_4$, sex (two levels); $X_5$, perform (two levels); and $X_6$, logarithm of Breslow thickness (in mm). Among all six covariates, $X_3$ and $X_6$ have missing values and the total number of completely observed samples is $n=234$. The data set is available from \cite{ibrahim2001bayesian}.

To illustrate the proposed method, we assume that the original data set fits into a logistic regression and we minimize the penalized pairwise pseudo likelihood (\ref{plike}) to obtain the estimates. In contrast, under the MAR assumption, the corresponding estimates can be calculated by a penalized logistic regression with the completely observed subjects. We examine both methods using three penalty functions: LASSO, SCAD and MCP. The variable selection and parameter estimation results are reported in Table \ref{table:data1}.

The comparison of the results shown by both methods is as follows. Variables sex, perform, log(Breslow) are never selected by any method or any penalty, showing some agreement of the two methods. However, variable age is selected by the proposed method but not the method assuming MAR; variable nodes1 is selected by either method and either penalty, but the proposed method always show an elevation of the parameter estimate; the selection of the variable treatment depends on the method and the penalty.

A similar data set was previously analyzed in \cite{ibrahim2001bayesian} and \cite{garcia2010variable}, and the latter showed that, both variable age and variable treatment can be selected by the adaptive LASSO method but not by the SCAD method. Variable age is negatively associated with a longer survival time, and its effect is not significant in the maximum likelihood estimate (MLE) method; while variable treatment is positively associated with a longer survival time, and its effect is significant according to the MLE.
Both these agreements and disagreements of these methods reveal some more information that is contained in the data but cannot be disclosed if only one single method is explored. This could certainly provide more insight of the data to investigators and clinicians.

\subsection{Prostate Cancer Study}

We also analyze a data set from a study (GEO GDS3289) investigating the association between prostate cancer tumors and genomic biomarkers \citep{tomlins2007integrative}. The whole data set can be accessed from the website of the National Center for Biotechnology Information of the National Institutes of Health. Briefly, this data set contains $N=104$ samples, out of which 34 are benign epithelium samples ($Y=0$) and 70 non-benign samples ($Y=1$).

There are missing values for various biomarkers in this data set. In our analysis, we include $p=64$ biomarkers in total and six of them have missing values with the number of missing samples for each biomarker ranging from 1 to 53.
The missing values result in a complete data set with the sample size $n=49$, and there are 36 non-benign samples in this complete data set. We adopt the penalized logistic regression in this analysis, and we examine the results under two different assumptions: one assuming MAR, and the other assuming (\ref{assume}), with three different representative penalty functions: LASSO, SCAD and MCP. Similar to the previous data analysis, the variable selection and the parameter estimation results are reported in Table \ref{table:data2}.

Our major findings and the comparison with previous literature can be summarized as follows.
First, some biomarkers like RHOB, can be selected by either method or either penalty function. Second, some other biomarkers, for example, MME, ANXA1, CLDN4 and SOX4 can be selected by our proposed method but not the method assuming MAR. Interestingly, they were all investigated in the previous literature \cite{kalin2011novel, geary2014caf, maeda2012claudin, wang2013sox4} and clinically concluded to be associated with the prostate cancer. Although we cannot reach a uniform conclusion that our method outperforms the MAR method in this real data exploration, the analysis demonstrates that it can reveal some extra genetic information by using our proposed method. This illustrates the potential usefulness of our proposed method and it will be very interesting to medical investigators and clinical practitioners.

\section{Discussion}

This paper addresses the problem of variable selection when missing values are present in the data set. Since the missing data mechanism assumptions are unverifiable, we adopt a very flexible and generally applicable one. The situations we consider include both ignorable and nonignorable missing data mechanisms. We allow both the number of nonzero components in the parameter $\bfbeta$ and the logarithm of the number of covariates to grow with the sample size. In particular, the logarithm of the number of covariates can grow at most as fast as an exponential rate of the sample size.

One may observe that, the proposed method only uses the information contained in the completely observed samples. In real applications, there may exist many partially observed samples, for example, maybe the covariate $\bmX$ values are always available. It will be difficult to directly get these partially observed samples involved in the current proposed approach. Some imputation techniques, for example \cite{chen2013variable, long2015variable, liu2016variable}, may be helpful and it warrants further study. Also, how to conduct the high dimensional statistical inference, especially the post-selection inference, is very interesting but challenging when the data contain missing values, which is beyond the scope of this paper and it certainly warrants further investigation.

Finally, we provide some practical guidance on using the proposed method. In reality, the missing data mechanism assumption is unverifiable and its underlying truth is unknown. Our major assumption (\ref{assume}) is more flexible than a single parametric assumption, and hence more generally applicable. From our real data analyses in Section 6, the proposed method and the method assuming MAR will always have some agreement and some disagreement. Although we can't reach a definite conclusion in reality, our proposed approach and analysis may provide some more insight on the real data, especially when the MAR assumption is in suspicion.

\section*{Appendix}


\begin{proof}[Verification of the Assumption 2]
  For the lower bound of $\rho_-(s)$, denote $F_{ij} = \{|y_i|\leq \tau\} \cap \{|y_j|\leq \tau\}$, where $\tau$ is a positive constant, we have
  \begin{eqnarray*}
  \nabla^2 \mathcal{L}(\bmgamma^\ast) &\geq& \frac{2}{n(n-1)}\sum_{1\leq i<j\leq n}\{\psi''(\yij\bmxij^T\bmgamma^\ast)\yij^2\bmxij^{\otimes 2}I(F_{ij})\} \\
  &\geq& c_3 \frac{2}{n(n-1)}\sum_{1\leq i<j\leq n}\{\yij^2\bmxij^{\otimes 2}I(F_{ij})\} \triangleq \matrW,
  \end{eqnarray*}
  where $c_3 = \exp(-4B\tau)\{1+\exp(4B\tau)\}^{-2}$.

According to the arguments in the proof of Theorem 3.10 in \cite{ning2017likelihood}, for any $\bmv \in \mathcal{F}$, where
\[
\mathcal{F} = \{\bmDelta \in R^p: \|\bmDelta\|_0=s, \|\bmDelta\|_2=1\},
\]
we have
\begin{eqnarray*}
  |\bmv^T\matrW\bmv - \bmv^T E(\matrW)\bmv| &\leq& \|\bmv\|_1^2 \|\matrW-E(\matrW)\|_\infty \\
  &\leq& s \|\matrW-E(\matrW)\|_\infty.
\end{eqnarray*}
Hence, $\rho_-(\matrW, s) \geq \rho_-(E(\matrW), s) - s \|\matrW-E(\matrW)\|_\infty$.
Note that the kernel function of $\matrW$ is bounded, i.e.,
$\|c_3 \yij^2 \bmxij^{\otimes 2} I(F_{ij})\|_\infty \leq 16 c_3 M^2 \tau^2$.
Then the Hoeffding's inequality can be applied to the centered U-statistics $W_{jk} - E(W_{jk})$. For some constant $t>0$ to be chosen, there exist some universal constants $c_4$, $c_5>0$, such that
\begin{eqnarray*}
\Pr\left( s \|\matrW-E(\matrW)\|_\infty > t \right) &\leq& \sum_{j,k} \Pr\left( |W_{jk}-E(W_{jk})|>\frac{t}{s} \right)\\
&\leq& c_4 p^2 \exp\left( -\frac{c_5 t^2 n}{s^2} \right).
\end{eqnarray*}

If $Y$ follows the normal linear model, without loss of generality, we assume $Y|\bmX \sim N(\alpha+\beta^T \bmX, \phi)$, then
\begin{eqnarray*}
  & & E\left( \yij^2 I(F_{ij}) | \bmx_i, \bmx_j \right)\\
  &=& \frac1{\sqrt{2\pi}}\int_{-\tau}^{\tau}\int_{-\tau}^{\tau} \yij^2 \exp\left\{ -\frac{(y_i-\alpha-\bmx_i^T \bmbeta)^2+(y_j-\alpha-\bmx_j^T \bmbeta)^2}{2\phi} \right\} dy_i dy_j\\
  &\geq& \frac1{\sqrt{2\pi}}\int_{-\tau}^{\tau}\int_{-\tau}^{\tau} \yij^2 \exp\left\{ -\frac{y_i^2+y_j^2+2B^2+2B|y_i|+2B|y_j|}{2\phi} \right\} dy_i dy_j\triangleq c_6.
\end{eqnarray*}
Therefore, we have
\begin{eqnarray*}
  \bmv^T E(\matrW) \bmv &=& \bmv^T E(E(\matrW|\bmx))\bmv \geq c_6 \bmv^T E\bmxij^{\otimes 2}\bmv\\
  &=& 2 c_6 \bmv^T E(\bmx_i\bmx_i^T)\bmv \geq 2 c_6 \lambda_{\min}(\Sigma_x)
\end{eqnarray*}
and hence, $\rho_-(E(\matrW), s) \geq 2 c_6 \lambda_{\min}(\Sigma_x)$, where $\Sigma_x = \Cov(\bmX)$.
By the Hoeffding equality, taking $t=c_6 \lambda_{\min}(\Sigma_x)$ we have
\[
\rho_-(s) \geq \rho_-(\matrW, s) \geq c_6 \lambda_{\min}(\Sigma_x),
\]
with probability at least $1-c_4 p^2 \exp(-c_5 c_6^2 \lambda^2_{\min}(\Sigma_x) n/s^2)$.

For the upper bound of $\rho_+(s)$, notice that
\[
\nabla^2\mathcal{L}(\bmgamma^\ast) \leq \frac{2}{n(n-1)}\sum_{1\leq i<j\leq n}\yij^2\bmxij^{\otimes 2} \triangleq \matrW'
\]
Similar as before, we have
\[
\rho_+(s) \leq \rho_+(\matrW', s) \leq \rho_+(E(\matrW'), s) + s \|\matrW'-E(\matrW')\|_\infty
\]

If $Y|\bmX \sim N(\alpha+\beta^T \bmX, \phi)$, we have
\[
E(\yij^2|\bmx_i, \bmx_j) = 2\phi + (\bmxij^T \bmbeta)^2
\]
and hence
\begin{eqnarray*}
  \rho_+(E(\matrW'), s) &\leq& E(2\phi (\bmxij^T \bmxij)^2) + E\{(\bmxij^T \bmbeta)^2 (\bmxij^T \bmv)^2\}\\
  &\leq& 4\phi \lambda_{\rm max}(\Sigma_x) + \frac12 E(\bmxij^T \bmbeta)^4 + \frac12 E(\bmxij^T \bmv)^4\\
  &\leq& 4\phi \lambda_{\rm max}(\Sigma_x) + 16B^4 + 16M^4.
\end{eqnarray*}
Following the similar argument as above, we have
\[
\rho_+(s) \leq \rho_+(\matrW', s) \leq t + 4\phi \lambda_{\rm max}(\Sigma_x) + 16B^4 + 16M^4
\]
with probability at least $1-c_1 p^2 \exp(-c_2 t^2 n/s^2)$, for any constant $t>0$.
For simplicity, after taking $t = c_6 \lambda_{\min}(\Sigma_x)$, we have
\[
\rho_+(s) \leq \rho_+(\matrW', s) \leq c_6 \lambda_{\min}(\Sigma_x) + 4\phi \lambda_{\rm max}(\Sigma_x) + 16B^4 + 16M^4
\]
with probability at least $1-c_4 p^2 \exp(-c_5 c_6^2 \lambda^2_{\min}(\Sigma_x) n/s^2)$.
The choices of $\rho_\ast$ and $\rho^\ast$ can be decided accordingly. When $Y|\bmX$ follows a logistic regression, based on the arguments in the proof of Theorem 3.10 in \cite{ning2017likelihood} and the above steps, the same conclusion follows. This completes the verification by taking $C_1 = 2 c_4$ and $C_2 = c_5 c_6^2 \lambda^2_{\min}(\Sigma_x)$.

\end{proof}

\begin{proof}[Proof of Lemma 3]
  First, by Lemma 1,
  \[
  \nabla\mathcal{L}(\bmgamma^\ast)=\frac{2}{n(n-1)}\sum_{1\leq i<j\leq n} \{\psi'(\yij \bmxij^T \bmgamma^\ast)\yij\bmxij - \yij\bmxij\}
  \]
  is a mean-zero U-statistic of order 2. Given the Assumption 1, we have
  \[
  \|\{\psi'(\yij\bmxij^T \bmgamma^\ast)\yij\bmxij - \yij\bmxij\}\|_{\infty} \leq 2M |\yij|.
  \]
  By the sub-exponential tail condition on $y_i$, for any $x>0$ and $u=1,\ldots,p$,
  \begin{eqnarray*}
    & & \Pr(|\{\psi'(\yij\bmxij^T \bmgamma^\ast)\yij\bmxij - \yij\bmxij\}_u|>x)\\
    &\leq& \Pr(|\yij|>x/(2M))\\
    &\leq& \Pr(|y_i|>x/(4M)) + \Pr(|y_i|>x/(4M))\\
    &\leq& 2c_1 \exp\{-c_2 x/(4M)\}.
  \end{eqnarray*}
  By Lemma 2 with $k=\lfloor n/2\rfloor$, we have
  \begin{eqnarray*}
    & & \Pr(\|\nabla\mathcal{L}(\bmgamma^\ast)\|_\infty > C_3\sqrt{\log p/n}) \leq \sum_{u=1}^p \Pr(|\nabla_u \mathcal{L}(\bmgamma^\ast)|>C_3\sqrt{\log p/n})\\
    &\leq& 2p \exp\left[ -\min \left\{ \frac{c_2^2 C_3^2 k \log p}{2^9 c_1^2 M^2 n}, \frac{c_2 C_3 k (\log p)^{1/2}}{2^5 c_1 M n^{1/2}} \right\} \right],
  \end{eqnarray*}
  which completes the proof by defining $C_4 = \frac{c_2^2 C_3^2}{3\cdot 2^9 c_1^2 M^2}$ and $C_5 = \frac{c_2 C_3}{3 \cdot 2^5 c_1 M }$, where we use the fact that $k/n>1/3$.
\end{proof}

\begin{proof}[Proof of Lemma 4]
We restrict all vectors on $S$ in this proof. For the sake of easy presentation, the subscript $S$ is omitted throughout.
From the Taylor's expansion, we have
\[
\what{\bmgamma}_{\rm O} - \bmgamma^\ast = -\{\nabla^2\mathcal{L}(\wtilde\bmgamma_1)\}^{-1}\nabla\mathcal{L}(\bmgamma^\ast),
\]
where $\wtilde\bmgamma_1 = \bmgamma^\ast + t_1(\what{\bmgamma}_{\rm O} - \bmgamma^\ast)$, $0\leq t_1\leq 1$.
Therefore
\[
\|\what{\bmgamma}_{\rm O} - \bmgamma^\ast\|_\infty \leq \|\{\nabla^2\mathcal{L}(\wtilde\bmgamma_1)\}^{-1}\|_{L_\infty} \|\nabla\mathcal{L}(\bmgamma^\ast)\|_\infty.
\]

For $\|\nabla\mathcal{L}(\bmgamma^\ast)\|_\infty$, based on the proof of Lemma 3, we have $\|\nabla\mathcal{L}(\bmgamma^\ast)\|_\infty \leq C_3\sqrt{\frac{\log s^\ast}{n}}$ with probability at least $1 - 2s^{\ast} \exp\left[ -\min \left\{ C_4\log s^{\ast}, C_5 n^{1/2} (\log s^{\ast})^{1/2} \right\} \right]$.

For $\|\{\nabla^2\mathcal{L}(\wtilde\bmgamma_1)\}^{-1}\|_{L_\infty}$, following the similar argument in \cite{ning2017likelihood}, we have
$\|\what{\bmgamma}_{\rm O} - \bmgamma^\ast\|_1 \leq c_{11} s^\ast \sqrt{\frac{\log s^\ast}{n}}$
with probability at least $1-c_{12} p^{-1}$ and
\[
\|\{\nabla^2\mathcal{L}(\bmgamma^\ast)\}^{-1}\{\nabla^2\mathcal{L}(\wtilde\bmgamma_1) - \nabla^2\mathcal{L}(\bmgamma^\ast)\}\|_{L_\infty} \leq s^\ast \min{\{e^b-1, 1-e^{-b}\}},
\]
where
\begin{eqnarray*}
b &=& \max_{i,j} |\yij\bmxij^T (\wtilde\bmgamma_1-\bmgamma^\ast)| \\
 &\leq&  \max_{i,j}  |Y_i-Y_j| \|\bmX_i-\bmX_j\|_\infty \|\wtilde\bmgamma_1 - \bmgamma^\ast\|_1 \\
 &\leq& 12c_2^{-1}\log(n)M c_{11} s^\ast \sqrt{\frac{\log s^\ast}{n}}
\end{eqnarray*}
with probability at least $1-c_1 n^{-1}-c_{12} p^{-1}$ by taking $\delta=3c_2^{-1}\log(n)$ defined in Assumption 1.
Therefore, $\|\{\nabla^2\mathcal{L}(\bmgamma^\ast)\}^{-1}\{\nabla^2\mathcal{L}(\wtilde\bmgamma_1) - \nabla^2\mathcal{L}(\bmgamma^\ast)\}\|_{L_\infty}$ is bounded by a term with the order of $\log(n) (s^\ast)^2 \sqrt{\frac{\log s^\ast}{n}} = o_p(1)$ with a high probability.
Then we can choose a sufficiently large $n$, such that
$\|\{\nabla^2\mathcal{L}(\bmgamma^\ast)\}^{-1}\{\nabla^2\mathcal{L}(\wtilde\bmgamma_1) - \nabla^2\mathcal{L}(\bmgamma^\ast)\}\|_{L_\infty} \leq 1/2$.
Then based on the Theorem 2.3.4 in \cite{golub1996matrix}, we have
\[
\|\nabla^2\mathcal{L}(\wtilde\bmgamma_1)^{-1}\|_{L_\infty} \leq \frac{\|\nabla^2\mathcal{L}(\bmgamma^\ast)^{-1}\|_{L_\infty}}{1-\|\{\nabla^2\mathcal{L}(\bmgamma^\ast)\}^{-1}\{\nabla^2\mathcal{L}(\wtilde\bmgamma_1) - \nabla^2\mathcal{L}(\bmgamma^\ast)\}\|_{L_\infty}} < 2C,
\]
and this completes the proof.
\end{proof}

\begin{proof}[Proof of Lemma 5]
According to the Assumption 3, since $q_\lambda(t)$ satisfies the Lipschitz continuity condition, we have
\[
-\zeta_-\|\bmgamma_2-\bmgamma_1\|^2 \leq (q'_\lambda(\bmgamma_2)-q'_\lambda(\bmgamma_1))^T(\bmgamma_2-\bmgamma_1) \leq -\zeta_+\|\bmgamma_2-\bmgamma_1\|^2,
\]
which implies that the convex function $-\mathcal{Q}(\bmgamma)$ satisfies
\[
(\nabla(-\mathcal{Q}_\lambda(\bmgamma_2)) - \nabla(-\mathcal{Q}_\lambda(\bmgamma_1)))^T (\bmgamma_2 - \bmgamma_1) \leq \zeta_- \|\bmgamma_2 - \bmgamma_1\|_2^2,
\]
and
\[
(\nabla(-\mathcal{Q}_\lambda(\bmgamma_2)) - \nabla(-\mathcal{Q}_\lambda(\bmgamma_1)))^T (\bmgamma_2 - \bmgamma_1) \geq \zeta_+ \|\bmgamma_2 - \bmgamma_1\|_2^2.
\]

According to Theorem 2.1.5 and Theorem 2.1.9 in \cite{nesterov2013introductory}, the above two expressions are equivalent definitions of strong smoothness and strong convexity respectively. In other words, $-\mathcal{Q}_\lambda(\bmgamma)$ satisfies
\[
-\mathcal{Q}_\lambda(\bmgamma_2) \leq -\mathcal{Q}_\lambda(\bmgamma_1) - \nabla \mathcal{Q}(\bmgamma_1)^T (\bmgamma_2 - \bmgamma_1) + \frac{\zeta_-}{2}\|\bmgamma_2 - \bmgamma_1\|_2^2,
\]
and
\[
-\mathcal{Q}_\lambda(\bmgamma_2) \geq -\mathcal{Q}_\lambda(\bmgamma_1) - \nabla \mathcal{Q}(\bmgamma_1)^T (\bmgamma_2 - \bmgamma_1) + \frac{\zeta_+}{2}\|\bmgamma_2 - \bmgamma_1\|_2^2.
\]

For our loss function $\mathcal{L}(\bmgamma)$, by Taylor's expansion and the mean value theorem, we have
\[
\mathcal{L}(\bmgamma_2) = \mathcal{L}(\bmgamma_1) + \nabla \mathcal{L}(\bmgamma_1)^T (\bmgamma_2 - \bmgamma_1) + \frac12 (\bmgamma_2 - \bmgamma_1)^T \nabla^2 \mathcal{L}(t\bmgamma_1 + (1-t)\bmgamma_2)(\bmgamma_2 - \bmgamma_1),
\]
where $0\leq t\leq 1$. Since we assume $\|(\bmgamma_2 - \bmgamma_1)_{\bar{S}}\|_0 \leq s^\ast$, which implies $\|\bmgamma_2 - \bmgamma_1\|_0 \leq 2s^\ast$. Therefore, by the definition of sparse eigenvalue, we have
\[
\frac{(\bmgamma_2 - \bmgamma_1)^T}{\|\bmgamma_2 - \bmgamma_1\|_2} \nabla^2\mathcal{L}(t\bmgamma_1 + (1-t)\bmgamma_2) \frac{(\bmgamma_2 - \bmgamma_1)}{\|\bmgamma_2 - \bmgamma_1\|_2} \in [\rho_-(\nabla^2 \mathcal{L}, 2s^\ast), \rho_+(\nabla^2 \mathcal{L}, 2s^\ast)].
\]
Plugging this into the RHS of the Taylor expansion, we have
\[
\mathcal{L}(\bmgamma_2) \geq \mathcal{L}(\bmgamma_1) + \nabla \mathcal{L}(\bmgamma_1)^T (\bmgamma_2-\bmgamma_1) + \frac{\rho_-(\nabla^2 \mathcal{L}, 2s^\ast)}{2}\|\bmgamma_2 - \bmgamma_1\|_2^2,
\]
and
\[
\mathcal{L}(\bmgamma_2) \leq \mathcal{L}(\bmgamma_1) + \nabla \mathcal{L}(\bmgamma_1)^T (\bmgamma_2-\bmgamma_1) + \frac{\rho_+(\nabla^2 \mathcal{L}, 2s^\ast)}{2}\|\bmgamma_2 - \bmgamma_1\|_2^2.
\]

Putting all of the above four inequalities together, we have
\[
\mathcal{\wtilde L}_\lambda(\bmgamma_2) \geq \mathcal{\wtilde L}_\lambda(\bmgamma_1) + \nabla \mathcal{\wtilde L}_\lambda(\bmgamma_1)^T (\bmgamma_2-\bmgamma_1) + \frac{\rho_-(\nabla^2\mathcal{L}, 2s^\ast) - \zeta_-}{2}\|\bmgamma_2-\bmgamma_1\|_2^2,
\]
and
\[
\mathcal{\wtilde L}_\lambda(\bmgamma_2) \leq \mathcal{\wtilde L}_\lambda(\bmgamma_1) + \nabla \mathcal{\wtilde L}_\lambda(\bmgamma_1)^T (\bmgamma_2-\bmgamma_1) + \frac{\rho_+(\nabla^2\mathcal{L}, 2s^\ast) - \zeta_+}{2}\|\bmgamma_2-\bmgamma_1\|_2^2.
\]
\end{proof}

\begin{proof}[Proof of Theorem 1]
From the Karush-Kuhn-Tucker condition, we have
\[
\nabla\mathcal{\wtilde L}_\lambda(\what\bmgamma) + \lambda \what\bmxi = 0,
\]
where $\what\bmxi \in \partial \|\what\bmgamma\|_1$ represents the subgradient, i.e., $\hat\xi_j=\mbox{sign}(\hat\gamma_j)$, if $\hat\gamma_j\neq 0$; $\hat\xi_j\in [-1,1]$ if $\hat\gamma_j=0$.
Next, we show that, there exists some $\bmxi_{\rm O} \in \partial \|\what\bmgamma_{\rm O}\|_1$, such that $\what\bmgamma_{\rm O}$ satisfies the exactly same condition as above
\[
\nabla\mathcal{\wtilde L}_\lambda(\what\bmgamma_{\rm O}) + \lambda\bmxi_{\rm O} = 0.
\]

For $j\in S$, by the condition of the weakest signal strength and the result of Lemma 4,
with probability at least $1- \delta_2$,
when $n$ is sufficiently large,
\begin{eqnarray}
|(\what\bmgamma_{\rm O})_j|\geq |\gamma^\ast_j| - \|\what\bmgamma_{\rm O}-\bmgamma^\ast\|_\infty \geq 2\nu - 2C C_3 \sqrt{\log s^\ast/n} > \nu,
\label{highplemma4}
\end{eqnarray}
then by the condition of the penalty function, we have
\[
(\nabla \mathcal{Q}_\lambda(\what\bmgamma_{\rm O}) + \lambda \bmxi_{\rm O})_j = (\nabla \mathcal{P}_\lambda (\what\bmgamma_{\rm O}))_j = p'_\lambda((\what\bmgamma_{\rm O})_j) = 0.
\]
For $j \in \bar{S}$, $(\what\bmgamma_{\rm O})_j=0$, so $(\nabla \mathcal{Q}_\lambda(\what\bmgamma_{\rm O}))_j=0$, therefore
\[
(\nabla\mathcal{\wtilde L}_\lambda(\what\bmgamma_{\rm O}) + \lambda \bmxi_{\rm O})_j = (\nabla\mathcal{L}(\what\bmgamma_{\rm O}) + \lambda \bmxi_{\rm O})_j,
\]
so we can define $(\bmxi_{\rm O})_j = (-\frac{\nabla\mathcal{L}(\what\bmgamma_{\rm O})}{\lambda})_j$.
Note that we choose $\lambda \asymp \sqrt{\log p/n}$, and from the proof of Lemma 3,
with probability at least $1 - \delta_1$,
\begin{eqnarray}
\|\nabla\mathcal{L}(\what\bmgamma_{\rm O})\|_\infty \leq C_3 \sqrt{\log p/n}.
\label{highplemma3}
\end{eqnarray}
So we have $\bmxi_{\rm O} \in [-1, 1]$,
and therefore we've found $\bmxi_{\rm O}$, such that $\bmxi_{\rm O}\in \partial \|\what\bmgamma_{\rm O}\|_1$, and $\nabla\mathcal{\wtilde L}_\lambda(\what\bmgamma_{\rm O}) + \lambda \bmxi_{\rm O}=0$,
with probability at least $1-\delta_1-\delta_2$, by (\ref{highplemma4}), (\ref{highplemma3}) and the fact that $P(A\bigcap B)\geq P(A)+P(B)-1$, where $A$ and $B$ are two arbitrary events.

Next, we show that $\|(\what\bmgamma-\what\bmgamma_{\rm O})_{\bar{S}}\|_0 \leq s^\ast$. Due to the analysis of the convergence properties based on the MM algorithm, presented in \cite{zou2008one}, we only need to prove this result in the $l$-th iteration, i.e., for $\what\bmgamma^{(l)}$. In the $l$-th iteration, we define $G^{(l)} = \{k: \gamma^\ast_k = 0, \hat\omega_k^{(l-1)}\geq p'_\lambda(c_8 \lambda), k=1,\ldots,p \}$, representing the covariates who are unimportant but heavily penalized. Its complement $\overline{G^{(l)}} = \{k: \gamma^\ast_k \neq 0, \mbox{ or } \hat\omega_k^{(l-1)}< p'_\lambda(c_8 \lambda), k=1,\ldots,p\}$. It's clear that $S \subset \overline{G^{(l)}}$. If we define $H := \overline{G^{(l)}} - S = \{k: \gamma^\ast_k = 0, \hat\omega_k^{(l-1)}< p'_\lambda(c_8 \lambda), k=1,\ldots,p\}$, it's also clear that $S$ and $H$ are disjoint. We are going to first show that $|\overline{G^{(l)}}|\leq 2 s^\ast$ by induction.

For $l=1$, because we have $\hat\omega_k^{(0)}=\lambda$, $\overline{G^{(1)}}=S$, hence $|\overline{G^{(1)}}|\leq s^\ast$. Now we assume that $|\overline{G^{(l)}}|\leq 2 s^\ast$ for some integer $l$ and our goal is to prove that $|\overline{G^{(l+1)}}|\leq 2 s^\ast$.

Suppose $\what\bmgamma^{(l)}$ is the solution in the $l$-th iteration, from the Karush-Kuhn-Tucker condition, we have
\[
\nabla \mathcal{L}(\what\bmgamma^{(l)}) + \what\bmomega^{(l-1)} \circ \bmxi^{(l)} = 0,
\]
where $\bmxi^{(l)} \in \partial \|\what\bmgamma^{(l)}\|_1$. In the following, we denote $\bmdelta = \what\bmgamma^{(l)} - \bmgamma^\ast$.
By the mean value theorem, we have
\[
\nabla\mathcal{L}(\what\bmgamma^{(l)}) - \nabla\mathcal{L}(\bmgamma^\ast) = \nabla^2 \mathcal{L}(\wtilde\bmgamma)\bmdelta,
\]
where $\wtilde\bmgamma = t\bmgamma^\ast+(1-t)\what\bmgamma^{(l)}$, which
implies
\[
0\leq \bmdelta^T \nabla^2\mathcal{L}(\wtilde\bmgamma)\bmdelta = -\bmdelta^T \what\bmomega^{(l-1)}\circ \bmxi^{(l)} - \nabla\mathcal{L}(\bmgamma^\ast)^T\bmdelta.
\]

For the second term, Holder's inequality implies
\[
\nabla\mathcal{L}(\bmgamma^\ast)^T\bmdelta \geq -\|\nabla\mathcal{L}(\bmgamma^\ast)\|_\infty \|\bmdelta\|_1.
\]
For the first term, also use Holder's inequality, we have
\begin{eqnarray*}
\bmdelta^T (\what\bmomega^{(l-1)}\circ \bmxi^{(l)}) &=& \bmdelta_S^T(\what\bmomega^{(l-1)}\circ \bmxi^{(l)})_S + |\bmdelta_H^T\what\bmomega^{(l-1)}_H| + |\bmdelta_G^T\what\bmomega^{(l-1)}_G| \\
&\geq& -\|\bmdelta_S\|_1 \|\what\bmomega^{(l-1)}_S\|_\infty + \|\bmdelta_H\|_1 \|\what\bmomega^{(l-1)}_H\|_{\rm min} + \|\bmdelta_G\|_1 \|\what\bmomega^{(l-1)}_G\|_{\rm min}.
\end{eqnarray*}
Combining these two inequalities, we have
\[
-\|\bmdelta_S\|_1 \|\what\bmomega^{(l-1)}_S\|_\infty + \|\bmdelta_H\|_1 \|\what\bmomega^{(l-1)}_H\|_{\rm min} + \|\bmdelta_G\|_1 \|\what\bmomega^{(l-1)}_G\|_{\rm min} - \|\nabla\mathcal{L}(\bmgamma^\ast)\|_\infty \|\bmdelta\|_1 \leq 0.
\]
Hence
\[
p'_\lambda(c_8 \lambda)\|\bmdelta_G\|_1 \leq \|\bmdelta_G\|_1 \|\what\bmomega^{(l-1)}_G\|_{\rm min} \leq \|\nabla\mathcal{L}(\bmgamma^\ast)\|_\infty \|\bmdelta\|_1 + \|\bmdelta_S\|_1 \|\what\bmomega^{(l-1)}_S\|_\infty.
\]

Therefore, we have
\[
\left[p'_\lambda(c_8 \lambda)-\|\nabla\mathcal{L}(\bmgamma^\ast)\|_\infty\right]\|\bmdelta_G\|_1 \leq \left[\|\what\bmomega^{(l-1)}_S\|_\infty + \|\nabla\mathcal{L}(\bmgamma^\ast)\|_\infty\right]\|\bmdelta_{\bar{G}}\|_1,
\]
which implies
\[
\|\bmdelta_G\|_1 \leq \frac{\|\what\bmomega^{(l-1)}_S\|_\infty+\|\nabla\mathcal{L}(\bmgamma^\ast)\|_\infty}{p'_\lambda(c_8 \lambda)-\|\nabla\mathcal{L}(\bmgamma^\ast)\|_\infty}\|\bmdelta_{\bar{G}}\|_1 \leq c_{13} \|\bmdelta_{\bar{G}}\|_1,
\]
which is equivalent to
\[
\|\what\bmgamma^{(l)}-\bmgamma^\ast\|_1 \leq (1+c_{13}) \|\what\bmgamma^{(l)}_{G^{(l)}}-\bmgamma^\ast_{G^{(l)}}\|_1.
\]
Similarly, we can also show that
\[
\|\what\bmgamma^{(l)}-\bmgamma^\ast\|_2 \leq (1+c_{13})\|\what\bmgamma^{(l)}_{I^{(l)}}-\bmgamma^\ast_{I^{(l)}}\|_2.
\]

Next, following the proof of Lemma A.3 in \cite{yang2014semiparametric}, based on the Assumption 2 and the condition that $s^\ast \sqrt{\frac{\log p}{n}}=o_p(1)$,
with probability at least $1-\delta_3$,
we can establish the following crude rates of convergence for $l\geq 1$:
\begin{eqnarray}
\|\what\bmgamma^{(l)}-\bmgamma^\ast\|_2 \leq c_{14}\rho_\ast^{-1}\sqrt{s^\ast}\lambda.
\label{highpass2}
\end{eqnarray}
By the concavity of $p_\lambda$, for any $k \in A := \overline{G^{(l+1)}} - S$, we have $|\hat\gamma^{(l)}_k| \geq c_8 \lambda$. Therefore we have
\[
\sqrt{|A|}\leq \|\what\bmgamma_A^{(l)}\|_2/(c_8\lambda) = \|\what\bmgamma_A^{(l)}-\bmgamma^\ast_A\|_2/(c_8\lambda)\leq c_{14}\rho_\ast^{-1}\sqrt{s^\ast}/c_8 \leq \sqrt{s^\ast},
\]
where the first inequality follows from $|A|\leq \sum_{k\in A}|\hat\gamma_k^{(l)}|^2/(c_8\lambda)^2$, and the last inequality follows from the appropriate choice of $c_{14}$ by the similar argument in \cite{yang2014semiparametric}. Note that this implies that $|\overline{G^{(l+1)}}|\leq 2s^\ast$. Therefore, by induction, $|\overline{G^{(l)}}|\leq 2s^\ast$ for any $l\geq 1$.
Then, from (\ref{highpass2}) we can follow the similar arguments in \cite{zhang2013multi, yang2014semiparametric} to conclude that $\|(\what\bmgamma-\what\bmgamma_{\rm O})_{\bar{S}}\|_0 \leq s^\ast$,
with probability at least $1-\delta_3$.

Next we are showing $\what\bmgamma = \what\bmgamma_{\rm O}$ when $n$ is sufficiently large. By Lemma 5,
it yields
\[
\mathcal{\wtilde L}_\lambda(\what\bmgamma) \geq \mathcal{\wtilde L}_\lambda(\what\bmgamma_{\rm O}) + \nabla \mathcal{\wtilde L}_\lambda(\what\bmgamma_{\rm O})^T (\what\bmgamma-\what\bmgamma_{\rm O}) + \frac{\rho_-(\nabla^2 \mathcal{L}, 2s^\ast) - \zeta_-}{2}\|\what\bmgamma-\what\bmgamma_{\rm O}\|_2^2,
\]
and
\[
\mathcal{\wtilde L}_\lambda(\what\bmgamma_{\rm O}) \geq \mathcal{\wtilde L}_\lambda(\what\bmgamma) + \nabla \mathcal{\wtilde L}_\lambda(\what\bmgamma)^T (\what\bmgamma_{\rm O}-\what\bmgamma) + \frac{\rho_-(\nabla^2 \mathcal{L}, 2s^\ast) - \zeta_-}{2}\|\what\bmgamma_{\rm O}-\what\bmgamma\|_2^2.
\]
By the convexity of $L_1$ norm, we have
\[
\lambda \|\what\bmgamma\|_1 \geq \lambda \|\what\bmgamma_{\rm O}\|_1 + \lambda (\what\bmgamma - \what\bmgamma_{\rm O})^T \bmxi_{\rm O},
\]
and
\[
\lambda \|\what\bmgamma_{\rm O}\|_1 \geq \lambda \|\what\bmgamma\|_1 + \lambda (\what\bmgamma_{\rm O} - \what\bmgamma)^T \what\bmxi.
\]

Adding the above four inequalities, we have
\[
0 \geq (\nabla \mathcal{\wtilde L}_\lambda(\what\bmgamma) + \lambda \what\bmxi)^T (\what\bmgamma_{\rm O} - \what\bmgamma) + (\nabla \mathcal{\wtilde L}_\lambda(\what\bmgamma_{\rm O}) + \lambda \bmxi_{\rm O})^T (\what\bmgamma - \what\bmgamma_{\rm O}) + (\rho_-(\nabla^2 \mathcal{L}, 2s^\ast) - \zeta_-)\|\what\bmgamma - \what\bmgamma_{\rm O}\|_2^2.
\]
Since $\nabla \mathcal{\wtilde L}_\lambda(\what\bmgamma) + \lambda \what\bmxi=0$, $\nabla \mathcal{\wtilde L}_\lambda(\what\bmgamma_{\rm O}) + \lambda \bmxi_{\rm O}=0$, $\rho_-(\nabla^2 \mathcal{L}, 2s^\ast) - \zeta_->0$, we must have $\what\bmgamma = \what\bmgamma_{\rm O}$, i.e., we conclude that $\what\bmgamma$ is the oracle estimator $\what\bmgamma_{\rm O}$. Also, since $\min_{j \in S}|(\what\bmgamma_{\rm O})_j|>0$ and the fact that $\mbox{supp}(\what\bmgamma_{\rm O}) \subset S$, we have
\[
\mbox{supp}(\what\bmgamma) = \mbox{supp}(\what\bmgamma_{\rm O}) = \mbox{supp}(\bmgamma^\ast),
\]
with probability at least $1-\delta_1-\delta_2-\delta_3$,
where this high probability comes from (\ref{highplemma4}), (\ref{highplemma3}), (\ref{highpass2}) in the process of this proof, and the fact that $P(A\bigcap B\bigcap C)\geq P(A) +P(B\bigcap C)-1 \geq P(A) +P(B) + P(C)-2$ where $A$, $B$ and $C$ are three arbitrary events,
and
this completes the proof.
\end{proof}

\section*{Acknowledgment}

Research reported in this publication was supported by the National Center for Advancing Translational Sciences of the National Institutes of Health under award Number UL1TR001412. The content is solely the responsibility of the authors and does not necessarily represent the official views of the NIH.
The authors thank the Editor, the Associate Editor and two anonymous referees for their constructive comments and insightful suggestions, which have led to a significantly improved paper.

\bibliographystyle{asa}
\bibliography{reference_ZYN}

\begin{thebibliography}{46}
\newcommand{\enquote}[1]{``#1''}
\expandafter\ifx\csname natexlab\endcsname\relax\def\natexlab#1{#1}\fi

\bibitem[{Bickel et~al.(2009)Bickel, Ritov, and
  Tsybakov}]{bickel2009simultaneous}
Bickel, P.~J., Ritov, Y., and Tsybakov, A.~B. (2009), \enquote{Simultaneous
  analysis of Lasso and Dantzig selector,} \textit{The Annals of Statistics},
  1705--1732.

\bibitem[{Chan(2013)}]{chan2013nuisance}
Chan, K. C.~G. (2013), \enquote{Nuisance parameter elimination for proportional
  likelihood ratio models with nonignorable missingness and random truncation,}
  \textit{Biometrika}, 100, 269--276.

\bibitem[{Chen and Wang(2013)}]{chen2013variable}
Chen, Q. and Wang, S. (2013), \enquote{Variable selection for multiply-imputed
  data with application to dioxin exposure study,} \textit{Statistics in
  Medicine}, 32, 3646--3659.

\bibitem[{Fan and Li(2001)}]{fan2001variable}
Fan, J. and Li, R. (2001), \enquote{Variable selection via nonconcave penalized
  likelihood and its oracle properties,} \textit{Journal of the American
  Statistical Association}, 96, 1348--1360.

\bibitem[{Fan and Lv(2010)}]{fan2010selective}
Fan, J. and Lv, J. (2010), \enquote{A selective overview of variable selection
  in high dimensional feature space,} \textit{Statistica Sinica}, 20, 101.

\bibitem[{Fan and Lv(2011)}]{fan2011nonconcave}
--- (2011), \enquote{Nonconcave penalized likelihood with NP-dimensionality,}
  \textit{IEEE Transactions on Information Theory}, 57, 5467--5484.

\bibitem[{Fan et~al.(2014)Fan, Xue, and Zou}]{fan2014strong}
Fan, J., Xue, L., and Zou, H. (2014), \enquote{Strong oracle optimality of
  folded concave penalized estimation,} \textit{The Annals of Statistics}, 42,
  819.

\bibitem[{Fan and Tang(2013)}]{fan2013tuning}
Fan, Y. and Tang, C.~Y. (2013), \enquote{Tuning parameter selection in high
  dimensional penalized likelihood,} \textit{Journal of the Royal Statistical
  Society: Series B (Statistical Methodology)}, 75, 531--552.

\bibitem[{Fang et~al.(2017)Fang, Zhao, and Shao}]{fang2017likelihood}
Fang, F., Zhao, J., and Shao, J. (2017), \enquote{Imputation-based adjusted
  score equations in generalized linear models with nonignorable missing
  covariate values,} \textit{Statistica Sinica}, to appear.

\bibitem[{Friedman et~al.(2010)Friedman, Hastie, and
  Tibshirani}]{friedman2010regularization}
Friedman, J., Hastie, T., and Tibshirani, R. (2010), \enquote{Regularization
  paths for generalized linear models via coordinate descent,} \textit{Journal
  of Statistical Software}, 33, 1.

\bibitem[{Garcia et~al.(2010)Garcia, Ibrahim, and Zhu}]{garcia2010variable}
Garcia, R.~I., Ibrahim, J.~G., and Zhu, H. (2010), \enquote{Variable selection
  for regression models with missing data,} \textit{Statistica Sinica}, 20,
  149--165.

\bibitem[{Geary et~al.(2014)Geary, Nash, Adisetiyo, Liang, Liao, Jeong, Zandi,
  and Roy-Burman}]{geary2014caf}
Geary, L.~A., Nash, K.~A., Adisetiyo, H., Liang, M., Liao, C.-P., Jeong, J.~H.,
  Zandi, E., and Roy-Burman, P. (2014), \enquote{CAF-secreted annexin A1
  induces prostate cancer cells to gain stem cell--like features,}
  \textit{Molecular Cancer Research}, 12, 607--621.

\bibitem[{Golub and Van~Loan(1996)}]{golub1996matrix}
Golub, G.~H. and Van~Loan, C.~F. (1996), \textit{Matrix Computations}, Johns
  Hopkins University Press, Baltimore, MD.

\bibitem[{Hunter and Lange(2004)}]{hunter2004tutorial}
Hunter, D.~R. and Lange, K. (2004), \enquote{A tutorial on MM algorithms,}
  \textit{The American Statistician}, 58, 30--37.

\bibitem[{Ibrahim et~al.(2001)Ibrahim, Chen, and Sinha}]{ibrahim2001bayesian}
Ibrahim, J.~G., Chen, M.-H., and Sinha, D. (2001), \textit{Bayesian Survival
  Analysis}, Springer.

\bibitem[{Ibrahim et~al.(1999)Ibrahim, Lipsitz, and Chen}]{ibrahim1999missing}
Ibrahim, J.~G., Lipsitz, S.~R., and Chen, M.-H. (1999), \enquote{Missing
  covariates in generalized linear models when the missing data mechanism is
  non-ignorable,} \textit{Journal of the Royal Statistical Society: Series B
  (Statistical Methodology)}, 61, 173--190.

\bibitem[{Ibrahim et~al.(2008)Ibrahim, Zhu, and Tang}]{ibrahim2008model}
Ibrahim, J.~G., Zhu, H., and Tang, N. (2008), \enquote{Model selection criteria
  for missing-data problems using the EM algorithm,} \textit{Journal of the
  American Statistical Association}, 1648--1658.

\bibitem[{Kalbfleisch(1978)}]{kalbfleisch1978likelihood}
Kalbfleisch, J.~D. (1978), \enquote{Likelihood methods and nonparametric
  tests,} \textit{Journal of the American Statistical Association}, 73,
  167--170.

\bibitem[{K{\"a}lin et~al.(2011)}]{kalin2011novel}
K{\"a}lin, M. et~al. (2011), \enquote{Novel prognostic markers in the serum of
  patients with castration-resistant prostate cancer derived from quantitative
  analysis of the pten conditional knockout mouse proteome,} \textit{European
  Urology}, 60, 1235--1243.

\bibitem[{Kim and Shao(2013)}]{kim2013statistical}
Kim, J.~K. and Shao, J. (2013), \textit{Statistical Methods for Handling
  Incomplete Data}, CRC Press.

\bibitem[{Kirkwood et~al.(1996)Kirkwood, Strawderman, Ernstoff, Smith, Borden,
  and Blum}]{kirkwood1996interferon}
Kirkwood, J.~M., Strawderman, M.~H., Ernstoff, M.~S., Smith, T.~J., Borden,
  E.~C., and Blum, R.~H. (1996), \enquote{Interferon alfa-2b adjuvant therapy
  of high-risk resected cutaneous melanoma: the Eastern Cooperative Oncology
  Group Trial EST 1684.} \textit{Journal of Clinical Oncology}, 14, 7--17.

\bibitem[{Liang and Qin(2000)}]{liang2000regression}
Liang, K.-Y. and Qin, J. (2000), \enquote{Regression analysis under
  non-standard situations: a pairwise pseudolikelihood approach,}
  \textit{Journal of the Royal Statistical Society. Series B, Statistical
  Methodology}, 773--786.

\bibitem[{Little and Rubin(2002)}]{little2002statistical}
Little, R.~J. and Rubin, D.~B. (2002), \textit{Statistical Analysis with
  Missing Data}, Wiley, 2nd ed.

\bibitem[{Liu et~al.(2016)Liu, Wang, Feng, and Melanie}]{liu2016variable}
Liu, Y., Wang, Y., Feng, Y., and Melanie, M.~W. (2016), \enquote{Variable
  selection and prediction with incomplete high-dimensional data,} \textit{The
  Annals of Applied Statistics}, 10, 418--450.

\bibitem[{Loh and Wainwright(2012)}]{loh2012high}
Loh, P.-L. and Wainwright, M.~J. (2012), \enquote{High-dimensional regression
  with noisy and missing data: Provable guarantees with nonconvexity,}
  \textit{The Annals of Statistics}, 40, 1637--1664.

\bibitem[{Long and Johnson(2015)}]{long2015variable}
Long, Q. and Johnson, B.~A. (2015), \enquote{Variable selection in the presence
  of missing data: resampling and imputation,} \textit{Biostatistics}, 16,
  596--610.

\bibitem[{Maeda et~al.(2012)Maeda, Murata, Chiba, Takasawa, Tanaka, Kojima,
  Masumori, Tsukamoto, and Sawada}]{maeda2012claudin}
Maeda, T., Murata, M., Chiba, H., Takasawa, A., Tanaka, S., Kojima, T.,
  Masumori, N., Tsukamoto, T., and Sawada, N. (2012),
  \enquote{Claudin-4-targeted therapy using clostridium perfringens enterotoxin
  for prostate cancer,} \textit{The Prostate}, 72, 351--360.

\bibitem[{McCullagh and Nelder(1989)}]{mccullagh1989generalized}
McCullagh, P. and Nelder, J.~A. (1989), \textit{Generalized Linear Models},
  Chapman \& Hall/CRC, 2nd ed.

\bibitem[{Nesterov(2013)}]{nesterov2013introductory}
Nesterov, Y. (2013), \textit{Introductory Lectures on Convex Optimization: A
  Basic Course}, vol.~87, Springer Science \& Business Media.

\bibitem[{Ning et~al.(2017)Ning, Zhao, and Liu}]{ning2017likelihood}
Ning, Y., Zhao, T., and Liu, H. (2017), \enquote{A likelihood ratio framework
  for high dimensional semiparametric regression,} \textit{The Annals of
  Statistics}, to appear.

\bibitem[{Qin et~al.(2008)Qin, Shao, and Zhang}]{qin2008efficient}
Qin, J., Shao, J., and Zhang, B. (2008), \enquote{Efficient and doubly robust
  imputation for covariate-dependent missing responses,} \textit{Journal of the
  American Statistical Association}, 103, 797--810.

\bibitem[{Robins and Ritov(1997)}]{robins1997toward}
Robins, J.~M. and Ritov, Y. (1997), \enquote{Toward a curse of dimensionality
  appropriate (CODA) asymptotic theory for semi-parametric models,}
  \textit{Statistics in Medicine}, 16, 285--319.

\bibitem[{Shao and Zhao(2013)}]{shao2013estimation}
Shao, J. and Zhao, J. (2013), \enquote{Estimation in longitudinal studies with
  nonignorable dropout,} \textit{Statistics and Its Interface}, 6, 303--313.

\bibitem[{Tang et~al.(2003)Tang, Little, and Raghunathan}]{tang2003analysis}
Tang, G., Little, R.~J., and Raghunathan, T.~E. (2003), \enquote{Analysis of
  multivariate missing data with nonignorable nonresponse,}
  \textit{Biometrika}, 90, 747--764.

\bibitem[{Tibshirani(1996)}]{tibshirani1996regression}
Tibshirani, R. (1996), \enquote{Regression shrinkage and selection via the
  lasso,} \textit{Journal of the Royal Statistical Society. Series B
  (Methodological)}, 267--288.

\bibitem[{Tomlins et~al.(2007)}]{tomlins2007integrative}
Tomlins, S.~A. et~al. (2007), \enquote{Integrative molecular concept modeling
  of prostate cancer progression,} \textit{Nature Genetics}, 39, 41--51.

\bibitem[{Wang et~al.(2013)Wang, Zhang, Yang, Chang, Qi, Zhou, and
  Han}]{wang2013sox4}
Wang, L., Zhang, J., Yang, X., Chang, Y., Qi, M., Zhou, Z., and Han, B. (2013),
  \enquote{SOX4 is associated with poor prognosis in prostate cancer and
  promotes epithelial--mesenchymal transition in vitro,} \textit{Prostate
  Cancer and Prostatic Diseases}, 16, 301--307.

\bibitem[{Wang et~al.(2014)Wang, Liu, and Zhang}]{wang2014optimal}
Wang, Z., Liu, H., and Zhang, T. (2014), \enquote{Optimal computational and
  statistical rates of convergence for sparse nonconvex learning problems,}
  \textit{The Annals of Statistics}, 42, 2164--2201.

\bibitem[{Yang et~al.(2014)Yang, Ning, and Liu}]{yang2014semiparametric}
Yang, Z., Ning, Y., and Liu, H. (2014), \enquote{On semiparametric exponential
  family graphical models,} \textit{arXiv preprint arXiv:1412.8697}.

\bibitem[{Zhang(2010)}]{zhang2010nearly}
Zhang, C.-H. (2010), \enquote{Nearly unbiased variable selection under minimax
  concave penalty,} \textit{The Annals of Statistics}, 894--942.

\bibitem[{Zhang(2013)}]{zhang2013multi}
Zhang, T. (2013), \enquote{Multi-stage convex relaxation for feature
  selection,} \textit{Bernoulli}, 19, 2277--2293.

\bibitem[{Zhao(2017)}]{zhao2017reducing}
Zhao, J. (2017), \enquote{Reducing bias for maximum approximate conditional
  likelihood estimator with general missing data mechanism,} \textit{Journal of
  Nonparametric Statistics}, 29, 577--593.

\bibitem[{Zhao and Shao(2015)}]{zhao2015semiparametric}
Zhao, J. and Shao, J. (2015), \enquote{Semiparametric pseudo-likelihoods in
  generalized linear models with nonignorable missing data,} \textit{Journal of
  the American Statistical Association}, 110, 1577--1590.

\bibitem[{Zhao and Shao(2017)}]{zhao2017approximate}
--- (2017), \enquote{Approximate conditional likelihood for generalized linear
  models with general missing data mechanism,} \textit{Journal of System
  Science and Complexity}, 30, 139--153.

\bibitem[{Zhao and Yu(2006)}]{zhao2006model}
Zhao, P. and Yu, B. (2006), \enquote{On model selection consistency of Lasso,}
  \textit{The Journal of Machine Learning Research}, 7, 2541--2563.

\bibitem[{Zou and Li(2008)}]{zou2008one}
Zou, H. and Li, R. (2008), \enquote{One-step sparse estimates in nonconcave
  penalized likelihood models,} \textit{The Annals of Statistics}, 36,
  1509--1566.

\end{thebibliography}

\newpage
\noindent
Jiwei Zhao\\
Department of Biostatistics, \\
State University of New York at Buffalo,\\
Buffalo, NY 14214, U.S.A.\\
\noindent
E-mail: zhaoj@buffalo.edu\\

\noindent
Yang Yang\\
Department of Biostatistics, \\
State University of New York at Buffalo,\\
Buffalo, NY 14214, U.S.A.\\
\noindent
E-mail: yyang39@buffalo.edu\\

\noindent
Yang Ning\\
Department of Statistical Science, \\
Cornell University,\\
Ithaca, NY 14853, U.S.A.\\
\noindent
E-mail: yn265@cornell.edu
\vskip -5cm

\newpage

\begin{table}
\centering
\caption{Mean and standard deviation (SD; in parentheses) of $\#$FP and $\#$FN in simulation settings (S1)--(S2). The proposed method is compared to two other methods: the method with no missing data, which uses all simulated data; and the method assuming MAR, which uses completely observed samples only.}
\begin{tabular}{llrrrrr}
\hline
 & \multirow{2}{*}{Method} & \multirow{2}{*}{Penalty} & \multicolumn{2}{c}{$\rho=0$} & \multicolumn{2}{c}{$\rho=0.5$}\\
 & & & \#FP & \#FN & \#FP & \#FN \\
\hline
\multirow{9}{*}{p=8} & \multirow{3}{20mm}{with no\\ missing data} & LASSO & 1.72 (1.57) & 0 (0) & 1.28 (1.39) & 0 (0) \\
 & & SCAD & 0.92 (1.36) & 0 (0) & 0.62 (1.06) & 0 (0) \\
 & & MCP & 0.73 (1.48) & 0 (0) & 0.45 (0.99) & 0.01 (0.10) \\
 \cline{2-7}
 & \multirow{3}{*}{MAR}  & LASSO & 2.50 (1.53) & 0 (0) & 1.76 (1.56) & 0 (0) \\
 & & SCAD & 1.30 (1.40) & 0 (0) & 0.93 (1.29) & 0.01 (0.10) \\
  & & MCP & 1.04 (1.59) & 0 (0) & 0.68 (1.29) & 0 (0) \\
 \cline{2-7}
 & \multirow{3}{*}{proposed}  & LASSO & 2.34 (1.39) & 0 (0) & 2.28 (1.33) & 0 (0) \\
 & & SCAD  & 0.98 (1.25) & 0 (0) & 0.98 (1.22) & 0.02 (0.14) \\
 & & MCP   & 0.78 (1.31) & 0 (0) & 0.63 (1.12) & 0.04 (0.20) \\
\hline
\multirow{9}{*}{p=200} & \multirow{3}{20mm}{with no\\ missing data} & LASSO & 12.45 (9.90) & 0 (0) & 8.88 (9.54) & 0 (0) \\
 & & SCAD  & 3.72 (3.22) & 0 (0) & 1.41 (1.28) & 0.02 (0.14) \\
 & & MCP   & 1.38 (2.10) & 0 (0) & 1.09 (1.14) & 0.01 (0.10) \\
  \cline{2-7}
 & \multirow{3}{*}{MAR} & LASSO & 14.06 (10.99) & 0.01 (0.10) & 9.89 (8.57) & 0 (0) \\
 & & SCAD  & 6.57 (6.92) & 0.01 (0.10) & 7.13 (6.60) & 0.14 (0.35) \\
 & & MCP   & 2.23 (3.59) & 0.05 (0.22) & 2.42 (2.81) & 0.19 (0.39) \\
  \cline{2-7}
 & \multirow{3}{*}{proposed}  & LASSO & 12.54 (6.83) & 0.01 (0.10) & 9.81 (5.84) & 0 (0) \\
 & & SCAD  & 3.91 (2.86) & 0.01 (0.10) & 4.35 (2.58) & 0.04 (0.20) \\
 & & MCP   & 1.96 (1.79) & 0.03 (0.17) & 2.71 (1.99) & 0.10 (0.30) \\
\hline
\end{tabular}
\label{table:linear}
\end{table}

\begin{table}
\centering
\caption{Mean and standard deviation (SD; in parentheses) of $\#$FP and $\#$FN in simulation settings (S3)--(S4). The proposed method is compared to two other methods: the method with no missing data, which uses all simulated data; and the method assuming MAR, which uses completely observed samples only.}
\begin{tabular}{llrrrrr}
\hline
 & \multirow{2}{*}{Method} & \multirow{2}{*}{Penalty} & \multicolumn{2}{c}{$\rho=0$} & \multicolumn{2}{c}{$\rho=0.5$}\\
 & & & \#FP & \#FN & \#FP & \#FN \\
\hline
\multirow{9}{*}{p=8} & \multirow{3}{20mm}{with no\\ missing data} & LASSO & 2.09 (1.13) & 0 (0) & 2.42 (1.12) & 0 (0) \\
 & & SCAD  & 0.76 (1.09) & 0 (0) & 0.64 (1.10) & 0 (0) \\
 & & MCP   & 0.52 (1.00) & 0 (0) & 0.40 (0.96) & 0 (0) \\
 \cline{2-7}
 & \multirow{3}{*}{MAR}  & LASSO & 2.72 (1.07) & 0 (0) & 2.56 (1.13) & 0 (0) \\
 & & SCAD  & 0.81 (0.92) & 0 (0) & 1.03 (1.34) & 0.01 (0.10) \\
  & & MCP   & 0.56 (1.07) & 0 (0) & 0.54 (1.01) & 0 (0) \\
 \cline{2-7}
 & \multirow{3}{*}{proposed}  & LASSO & 2.32 (1.16) & 0 (0) & 2.47 (1.10) & 0 (0) \\
 & & SCAD  & 0.78 (1.08) & 0 (0) & 0.66 (1.09) & 0.01 (0.10) \\
 & & MCP   & 0.65 (1.12) & 0 (0) & 0.58 (1.12) & 0.01 (0.10) \\
\hline
\multirow{9}{*}{p=200} & \multirow{3}{20mm}{with no\\ missing data} & LASSO & 20.16 (10.93) & 0 (0) & 19.94 (8.28) & 0.04 (0.20) \\
 & & SCAD  & 9.77 (8.15) & 0 (0) & 14.08 (9.40) & 0.02 (0.20) \\
 & & MCP   & 2.28 (2.83) & 0 (0) & 3.78 (3.95) & 0.05 (0.26) \\
  \cline{2-7}
 & \multirow{3}{*}{MAR} & LASSO & 27.91 (14.38) & 0 (0) & 27.65 (14.95) & 0.46 (0.56) \\
 & & SCAD  & 16.87 (8.02) & 0.01 (0.10) & 18.77 (8.85) & 0.22 (0.48) \\
 & & MCP   & 4.33 (3.31) & 0.01 (0.10) & 5.28 (4.28) & 0.47 (0.70) \\
  \cline{2-7}
 & \multirow{3}{*}{proposed}  & LASSO & 23.09 (13.16) & 0 (0) & 24.62 (13.96) & 0.39 (0.53) \\
 & & SCAD  & 13.73 (5.88) & 0 (0) & 15.59 (6.76) & 0.14 (0.38) \\
 & & MCP   & 4.84 (3.08) & 0.02 (0.14) & 4.85 (3.49) & 0.22 (0.50) \\
\hline
\end{tabular}
\label{table:logistic}
\end{table}

\begin{table}
\centering
\caption{Mean and standard deviation (SD; in parentheses) of $\#$FP and $\#$FN in simulation settings (S5)--(S6). The proposed method is compared to two other methods: the method with no missing data, which uses all simulated data; and the method assuming MAR, which uses completely observed samples only.}
\begin{tabular}{llrrrrr}
\hline
 & \multirow{2}{*}{Method} & \multirow{2}{*}{Penalty} & \multicolumn{2}{c}{$\rho=0$} & \multicolumn{2}{c}{$\rho=0.5$}\\
 & & & \#FP & \#FN & \#FP & \#FN \\
\hline
\multirow{9}{*}{p=8} & \multirow{3}{20mm}{with no\\ missing data} & LASSO & 1.42 (1.44) & 0 (0) & 1.32 (1.48) & 0 (0) \\
 & & SCAD  & 0.62 (1.10) & 0 (0) & 1.10 (1.62) & 0 (0) \\
 & & MCP   & 0.46 (1.14) & 0 (0) & 0.65 (1.27) & 0 (0) \\
 \cline{2-7}
 & \multirow{3}{*}{MAR}  & LASSO & 2.19 (1.48) & 0 (0) & 1.90 (1.49) & 0.01 (0.10) \\
 & & SCAD  & 0.87 (1.16) & 0 (0) & 1.15 (1.29) & 0.03 (0.17) \\
  & & MCP   & 0.65 (1.13) & 0 (0) & 0.77 (1.18) & 0.02 (0.14) \\
 \cline{2-7}
 & \multirow{3}{*}{proposed}  & LASSO & 1.98 (1.21) & 0 (0) & 2.48 (1.38) & 0 (0) \\
 & & SCAD  & 0.79 (1.17) & 0 (0) & 1.20 (1.40) & 0.01 (0.10) \\
 & & MCP   & 0.58 (1.11) & 0 (0) & 0.81 (1.25) & 0.01 (0.10) \\
\hline
\multirow{9}{*}{p=200} & \multirow{3}{20mm}{with no\\ missing data} & LASSO & 11.20 (9.45) & 0 (0) & 8.12 (8.68) & 0 (0) \\
 & & SCAD  & 3.37 (3.41) & 0 (0) & 1.37 (1.59) & 0.02 (0.14) \\
 & & MCP   & 1.37 (2.30) & 0 (0) & 1.01 (1.31) & 0 (0) \\
  \cline{2-7}
 & \multirow{3}{*}{MAR} & LASSO & 14.00 (11.95) & 0.01 (0.10) & 7.91 (7.80) & 0 (0) \\
 & & SCAD  & 5.35 (5.29) & 0.02 (0.14) & 5.72 (5.53) & 0.10 (0.30) \\
 & & MCP   & 2.47 (3.41) & 0.06 (0.24) & 2.35 (2.92) & 0.17 (0.38) \\
  \cline{2-7}
 & \multirow{3}{*}{proposed}  & LASSO & 11.11 (5.96) & 0 (0) & 9.68 (6.59) & 0 (0) \\
 & & SCAD  & 3.67 (2.85) & 0.01 (0.10) & 4.02 (2.27) & 0.05 (0.22) \\
 & & MCP   & 2.10 (2.30) & 0.03 (0.17) & 2.46 (1.94) & 0.10 (0.30) \\
\hline
\end{tabular}
\label{table:linear2}
\end{table}

\begin{table}
\centering
\caption{Mean and standard deviation (SD; in parentheses) of computing time (in seconds) in simulation settings (S1)--(S2).}
\begin{tabular}{llcrrr}
\hline
 & & Method & LASSO & SCAD & MCP \\
\hline
\multirow{6}{*}{p=8} & \multirow{3}{*}{$\rho=0$} &  with no missing data & 0.04(0.00) & 0.02(0.00) & 0.02(0.00)\\
 & & MAR & 0.04(0.01) & 0.02(0.00) & 0.02(0.00)\\
 & & proposed & 0.45(0.06) & 1.47(0.76) & 1.05(0.15)\\
 \cline{2-6}
 & \multirow{3}{*}{$\rho=0.5$} &  with no missing data & 0.04(0.01) & 0.02(0.00) & 0.02(0.01)\\
 & & MAR & 0.04(0.02) & 0.02(0.00) & 0.02(0.00)\\
 & & proposed & 0.44(0.06) & 1.53(0.62) & 1.48(0.22)\\
\hline
\multirow{6}{*}{p=200} & \multirow{3}{*}{$\rho=0$} & with no missing data & 0.17(0.02) & 0.05(0.00) & 0.06(0.01)  \\
 & & MAR & 0.08(0.01) & 0.05(0.01) & 0.05(0.01)  \\
 & & proposed & 9.74(1.89) & 47.71(9.24) & 22.44(4.36)\\
  \cline{2-6}
& \multirow{3}{*}{$\rho=0.5$} & with no missing data & 0.17(0.02) & 0.05(0.01) & 0.05(0.01) \\
 & & MAR & 0.07(0.01) & 0.04(0.01) & 0.05(0.01) \\
 & & proposed  & 6.91(1.41) & 36.81(7.07) & 19.58(3.82) \\
\hline
\end{tabular}
\label{table:time}
\end{table}

\begin{table}
\centering
\caption{The variable selection and parameter estimation results in the melanoma study contrasting the method assuming MAR and the proposed method.}
\begin{tabular}{lrrrrrr}
\hline
& \multicolumn{2}{c}{LASSO} & \multicolumn{2}{c}{SCAD} & \multicolumn{2}{c}{MCP} \\
\cline{2-7}
& MAR & proposed & MAR & proposed & MAR & proposed \\
\hline
$|\{i: \hat\gamma_i\neq 0\}|$ & 2 & 3 & 2 & 2 & 1 & 2 \\
\hline
treatment & -0.035 & -0.024 & -0.022 & 0.000 & 0.000 & 0.000 \\
age & 0.000 & 0.014 & 0.000 & 0.016 & 0.000 & 0.016 \\
nodes1 & 0.422 & 0.564 & 0.539 & 0.691 & 0.528 & 0.691 \\
sex & 0.000 & 0.000 & 0.000 & 0.000 & 0.000 & 0.000\\
perform & 0.000 & 0.000 & 0.000 & 0.000 & 0.000 & 0.000\\
log(Breslow) & 0.000 & 0.000 & 0.000 & 0.000 & 0.000 & 0.000\\
\hline
\end{tabular}
\label{table:data1}
\end{table}

\begin{table}
\centering
\caption{The variable selection and parameter estimation results in the prostate cancer study contrasting the method assuming MAR and the proposed method.}
\begin{tabular}{lrrrrrr}
\hline
& \multicolumn{2}{c}{LASSO} & \multicolumn{2}{c}{SCAD} & \multicolumn{2}{c}{MCP} \\
\cline{2-7}
& MAR & proposed & MAR & proposed & MAR & proposed \\
\hline
$|\{i: \hat\gamma_i\neq 0\}|$ & 13 & 13 & 10 & 11 & 3 & 6 \\
\hline
  RHOB & -4.593 & -3.224 & -3.640 & -0.868 & -2.459 & -0.885 \\
  MME & -0.111 & -0.821 & 0.000 & -0.276 & 0.000 & -0.588 \\
  ANXA1 & 0.000 & -0.280 & 0.000 & -0.950 & -1.140 & -0.690 \\
  FAM89A & -2.917 & -2.574 & -0.937 & -0.779 & 0.000 & 0.000 \\
  SETD5 & 1.279 & 2.727 & 0.555 & 0.434 & 0.000 & 0.000 \\
  CLDN4 & 0.000 & 0.577 & 0.000 & 0.483 & 0.000 & 0.586 \\
  SOX4 & 0.000 & 3.352 & 0.000 & 1.199 & 0.000 & 2.016 \\
  IMAGE:133130 & 0.000 & 2.455 & 0.000 & 1.418 & 0.000 & 3.487 \\
  ADAM22 & -1.240 & 0.000 & -0.535 & 0.000 & -6.052 & 0.000 \\
  AMACR & 0.098 & 1.066 & 0.000 & 0.345 & 0.000 & 0.000 \\
  ODF2 & 3.039 & 0.042 & 0.519 & 0.000 & 0.000 & 0.000 \\
  ST14 & 2.513 & 0.490 & 0.843 & 0.000 & 0.000 & 0.000 \\
  IMAGE:490971 & 0.638 & 0.000 & 1.615 & 0.000 & 0.000 & 0.000 \\
  RND3 & -2.251 & 0.000 & -0.182 & 0.000 & 0.000 & 0.000 \\
  KIAA0020 & 8.011 & 0.000 & 4.559 & 0.000 & 0.000 & 0.000 \\
  SLC25A6 & 1.660 & 0.000 & 0.485 & 0.000 & 0.000 & 0.000 \\
  MYO6 & 0.000 & 0.965 & 0.000 & 0.385 & 0.000 & 0.000 \\
  MYC & 0.000 & 0.117 & 0.000 & 0.000 & 0.000 & 0.000 \\
  EFEMP2 & -0.977 & 0.000 & 0.000 & 0.000 & 0.000 & 0.000 \\
  SERPING1 & 0.000 & 0.000 & 0.000 & -0.163 & 0.000 & 0.000 \\
  others & 0.000 & 0.000 & 0.000 & 0.000 & 0.000 & 0.000 \\
\hline
\end{tabular}
\label{table:data2}
\end{table}

\end{document}